\documentclass[a4paper,11pt]{article}
\pdfoutput=1 

\usepackage{jheppub} 
\usepackage[T1]{fontenc} 
\usepackage{xcolor}
\usepackage{subfigure}
\usepackage{comment}
\usepackage[export]{adjustbox}
\usepackage{amsmath}
\usepackage{bbold}

\newcommand{\ket}[1]{\left|#1\right\rangle}
\newcommand{\bra}[1]{\left\langle #1\right|}
\newcommand{\tr}{\text{Tr}}

\newcommand{\be}{\begin{equation}}
\newcommand{\ee}{\end{equation}}
\newcommand\vep{\varepsilon}

\newcommand{\PG}[1]{\textcolor{magenta}{#1}}
\newcommand{\fontH}[1]{#1}
\newcommand{\fonth}[1]{#1}

\def\text#1{{\rm #1}}

\def\avg#1{\left\langle#1\right\rangle}
\def\bra#1{\left\langle#1\right|}
\def\ket#1{\left|#1\right\rangle}
\def\braket#1#2{\left\langle #1\right|\left.#2\right\rangle}

\def\kc#1{\left(#1\right)}
\def\kd#1{\left[#1\right]}

\def\be{\begin{equation}}       \def\ee{\end{equation}}
\def\bea{\begin{eqnarray}}      \def\eea{\end{eqnarray}}
\def\ba{\begin{array} }
\def\ea{\end{array} }

\def\=>{\Rightarrow}
\def\>{\rightarrow}

\newcommand{\norm}[1]{\left\lVert#1\right\rVert}

\newcommand{\rhoAW}{\rho_{\fonth A  \fonth W}}
\newcommand{\fA}{\fonth{A}}
\newcommand{\fB}{\fonth{B}}
\newcommand{\fC}{\fonth{C}}
\newcommand{\fL}{\fonth{L}}

\newcommand{\fW}{\fonth{W}}
\newcommand{\fE}{\fonth{E}}

\newtheorem{theorem}{Theorem}
\newcounter{facts}[section]

\renewcommand{\thefacts}{\arabic{facts}}

\title{\boldmath Space-time generalization of mutual information}

\author[1,2]{Paolo Glorioso,}
\author[1]{Xiao-Liang Qi}
\affiliation[1]{Stanford Institute for Theoretical Physics, Stanford University}
\author[1,3]{and Zhenbin Yang}
\affiliation[2]{Department of Physics \& Condensed Matter Theory Center, University of Maryland, College Park}
\affiliation[3]{Institute for Advanced Study, Tsinghua University}

\abstract{
The mutual information characterizes correlations between spatially separated regions of a system. Yet, in experiments we often measure dynamical correlations, which involve probing operators that are also separated in time. Here, we introduce a space-time generalization of mutual information which, by construction, satisfies several natural properties of the mutual information and at the same time characterizes correlations across subsystems that are separated in time. In particular, this quantity, that we call the \emph{space-time mutual information}, bounds all dynamical correlations. We construct this quantity based on the idea of the quantum hypothesis testing. As a by-product, our definition provides a transparent interpretation in terms of an experimentally accessible setup. We draw connections with other notions in quantum information theory, such as quantum channel discrimination. Finally, we study the behavior of the space-time mutual information in several settings and contrast its long-time behavior in many-body localizing and thermalizing systems.
}

\begin{document} 
\maketitle

\flushbottom
\section{Introduction}


A prominent experimental diagnostic for the properties of physical systems are correlation functions. These offer a window into the complex dependencies that exist among different subregions in space and time of a given system, and provide an organized and transparent way to characterize its fundamental properties. For example, the electric conductivity is determined by the current-current response function. Angle-resolved photo emission experiments measure the two-point function of the electron as a function of momentum and energy. Phases of matter are also characterized by correlation functions: for example, spontaneously broken symmetry phases, such as ferromagnetism and crystalline order, correspond to long-range correlations in space. In general, a two-point correlation function is defined for two operators each residing in a region of spacetime. 

One question that naturally arises is whether there exists a quantum information measure to quantify the amount of correlations between two space-time regions. For two spatial regions, $A$ and $B$, defined at the same time, mutual information $I(A:B)=S_A+S_B-S_{AB}$ is a natural measure, where $S_A$ represents the von Neumann entropy of subsystem $A$. $I(A:B) = 0$ if and only if the states of $AB$ factorize as $\rho_{AB}=\rho_A\otimes \rho_B$, which implies that there is no connected correlation between the two regions. A non-zero $I(A:B)$ provides a quantitative measure of correlations that is independent of the specific operators that are correlated. For instance, if we consider two ferromagnetic states, one with long-range correlations of the spin-$z$ component $\left\langle S_z(r)S_z(r')\right\rangle_c$ and another with an equal amount of correlation in $S_x$, then, keeping everything else fixed, both states will yield the same mutual information. Importantly, the mutual information provides an upper bound on the connected correlation functions between the two regions \cite{wolf2008area}.

In physical experiments, dynamical correlation functions can be measured for regions located at different times, just like equal-time correlators. However, using mutual information to measure correlations between different times is not feasible.
In quantum field theory, the mutual information can typically be defined for algebras associated with space-like separated regions A and B that are not adjacent, but it is likely undefined for time-like separated regions, particularly when region B is inside the future lightcone of region A. For finite dimensional quantum systems, the issue arises because the state $\rho_{AB}$ is typically not defined for time-like separated regions. Our study is motivated by the idea that a physical system should be characterized by observables. Thus, we aim to generalize mutual information for a pair of space-time regions A and B that is valid even when they are not space-like separated, and hence when $\rho_{AB}$ is not defined. This is the primary objective of our research.

In this paper, we introduce a novel quantity, called the space-time mutual information (STMI) denoted as $J(\fA:\fB)$, which generalizes the mutual information for two arbitrary space-time regions $\fA$ and $\fB$. Our approach is based on the idea of hypothesis testing where, in a gedanken experiment, an experimentalist has access to a physical system at regions $A$ and $B$ defined at separate times. For example, in a qubit chain with qubits labelled by $x=1,2,...,N$, region $A$ could be qubit $x=1$ at time $t=t_1$ and $B$ could be the qubits $x=2,3$ at time $t=t_2$. The experimentalist is allowed to couple a general ancilla to regions $A$ and $B$ of the system. 
This can include ordinary measurements as well as more sophisticated quantum couplings with the ancilla, such as applying a quantum perturbation at $A$ and measure its consequence in region $B$. The goal of the experimentalist is to distinguish between two situations. The first situation is where the experimentalist accesses regions $A$ and $B$ of the \emph{same} system, and therefore can measure correlations between them. The second situation where the coupling at $A$ and $B$ occurs in two \emph{independent} copies of the original system. By construction, in the second situation there is no correlation between the two subsystems. The difficulty of distinguishing the two situations measures the amount of correlation, which can be characterized by a relative entropy. The STMI is defined by optimizing this relative entropy over all possible system-ancilla coupling schemes. The advantage of our quantity is that it directly describes an experimentally accessible setup. 

The remainder of the paper is organized as follows. In Sec. \ref{sec:defi} we introduce the intuition and the definition of the STMI, and explore some of its simple properties. We show that the STMI reduces to the ordinary mutual information when regions $A$ and $B$ are spacelike separated. We investigate various properties of STMI such as its monotonicity under local operations at $A$ and $B$ separately. In Sec. \ref{sec:bound} we prove that the STMI is an upper-bound of connected \emph{dynamical} correlation functions, a direct generalization of the corresponding inequality in Ref. \cite{wolf2008area} for static correlations. In Sec. \ref{sec:conditional STMI} we show that, in certain cases, when three subregions are considered, our quantity satisfies a space-time generalization of the Markov property. In Sec. \ref{sec:qcd} we draw a connection between the STMI and quantum channel discrimination and use this to prove additivity of the STMI in a special setting. Sec. \ref{sec:ansatz} proposes a simplification of the definition of the STMI, which holds in a restricted regime and which we apply to provide semi-analytic results when studying the examples of Sec. \ref{sec:examples}. Finally, we introduce a classical counterpart of the STMI in Sec. \ref{sec:class} and discuss conclusions and outlook in Sec. \ref{sec:concl}.

\section{Definition of space-time mutual information}\label{sec:defi}

\subsection{General intuition}


\begin{figure}
    \centering
    \includegraphics[width=5in]{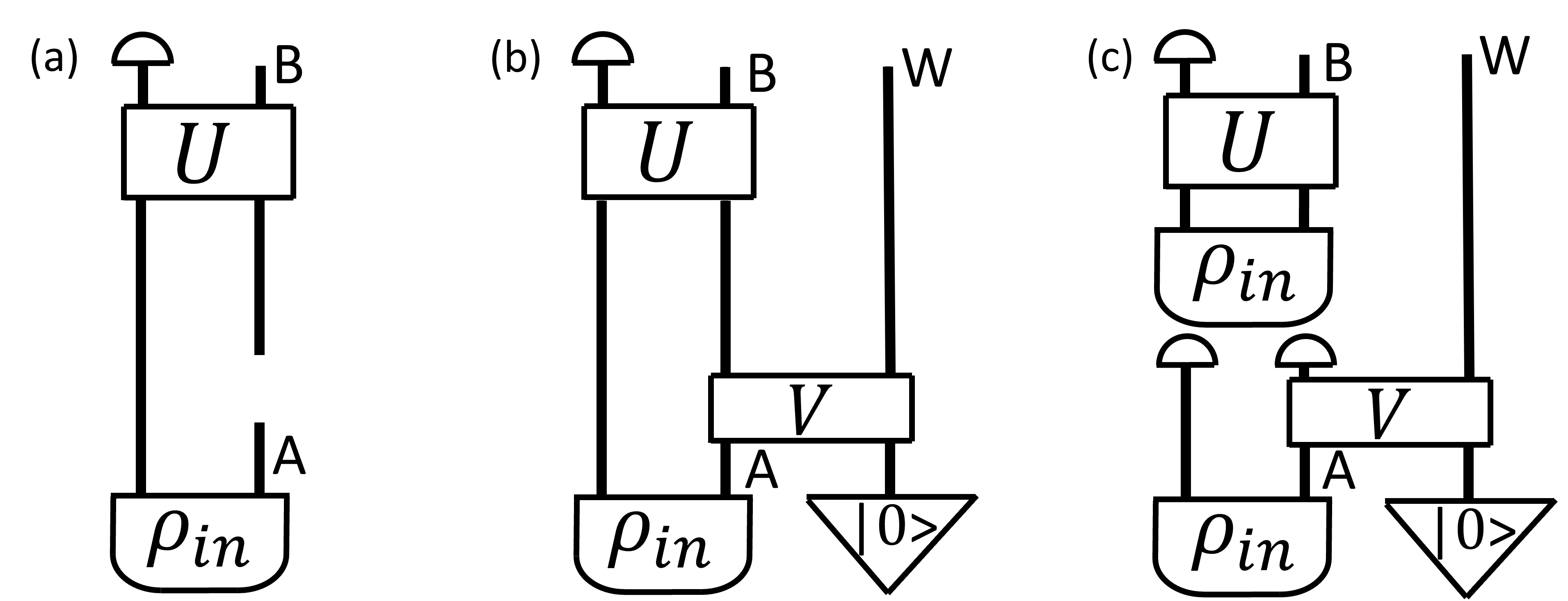}
    \caption{(a) System with initial state $\rho_{\text{in}}$ undergoing evolution with unitary $U$. The upper cap stands for tracing over the corresponding region. (b) Coupling between subsystem $\fA$ and ancilla $\fW$ giving rise to the connected state $\rho_{\fB\fW}$. (c) Disconnected state $\rho_{\fB,0}\otimes \rho_{\fW}$, where $\rho_{\fB,0}$ is the unperturbed evolved state reduced to subsystem $\fB$. }
    \label{fig:int1}
\end{figure}

Before presenting the rigorous definition of STMI, we would like to provide some intuition by presenting a simplified version of this quantity. Let us consider a system in an initial state $\rho_{\text{in}}$ defined on a Hilbert space on region $\fA\bar\fA$ that evolves by time evolution $U$ into an output state $\rho_{\text{out}}\equiv U\rho_{\text{in}}U^\dagger$, which can be partitioned into region $\fB$ and  $\bar\fB$. Here, $\fA$ and $\fB$ denote two subregions before and after the evolution, respectively, as in Fig. \ref{fig:int1}(a). We would like to introduce a generalization of mutual information that is applicable to general subregions $\fA$ and $\fB$, in particular to when these are causally connected. In the latter case the standard definition of mutual information does not apply, as there is no joint state on $\fA \fB$. To overcome this issue we couple subsystem $\fA$ to an ancilla $\fW$, which plays the role of an idler, allowing us to ``carry to the future'' the information encoded in $\fA$, as in Fig. \ref{fig:int1}(b). We shall denote the resulting state as $\rho_{\fB\fW}$. The operator $V$ that couples ancilla and system, as well as the dimension of $\fW$, are for now arbitrary. Without loss of generality, as we will see momentarily, we can always assume $V$ to be unitary, and the initial state of $\fW$ to be pure.  Now, recall that, when $\fA$ and $\fB$ are spatially separated and can be embedded in the same Hilbert space, the standard mutual information is the relative entropy between the connected state reduced on $\fA\fB$, $\rho_{\fA\fB}$, and the disconnected state $\rho_{\fA}\otimes\rho_{\fB}$. By analogy with this, in the case when $\fA$ and $\fB$ are causally connected, we consider the relative entropy between the connected state $\rho_{\fB\fW}$ and an analog of the disconnected state $\rho_{\fA}\otimes\rho_{\fB}$.
This disconnected state should be such that subregion $\fB$ is unaffected by the presence of the perturbation $V$ acting on $\fA$. The natural choice for such state is then given in Fig. \ref{fig:int1}(c), which we write as $\rho_{\fB,0}\otimes \rho_{\fW}$, where $\rho_{\fB,0}$ denotes the unperturbed evolved state in $\fB$, and $\rho_{\fW}={\rm tr}_{\fB}\left(\rho_{\fB\fW}\right)$ is the state of $W$ after coupling with $A$, which is determined by $\rho_{\text{in}}$ and coupling $V$ (which is therefore independent from the time evolution $U$). Intuitively, the state $\rho_{\fB,0}\otimes \rho_{\fW}$ is the state that determines the disconnected term $\left\langle O_B\right\rangle\left\langle O_A\right\rangle$ in correlation functions, for any operators $O_A,O_B$.

Finally, we define the space-time mutual information $ J_1(\fA:\fB)$ by maximizing the relative entropy over the ancilla-system coupling $V$:
\be\label{JAB} J_1(\fA:\fB)=\sup_V S(\rho_{\fB\fW}|\rho_{\fB,0}\otimes\rho_{\fW})\,.\ee
We now see that it is sufficient to consider unitary coupling between system and ancilla. Indeed, if we take $V$ to be a generic quantum channel, this is equivalent to having a unitary coupling to a bigger $\fontH{W}$ followed by partial trace, which will only reduce the relative entropy, so for the purpose of taking the supremum it is sufficient to consider unitaries. Since we assume $\fontH{W}$ is arbitrarily large, we can also take the initial state of $\fontH{W}$ to be a pure state. If the initial state is a mixed state, we can purify it by enlarging $\fontH{W}$. 

As we will show below, this definition already satisfies two important requirements: it reduces to the standard mutual information when $\fA$ and $\fB$ are spatially separated, and it bounds all space-time correlation functions of operators supported on $\fA$ and $\fB$, with any normal (i.e., defined on the Schwinger-Keldysh time contour) time ordering. 

The general definition of STMI is similar to Eq. (\ref{JAB}) except that it is defined for $N$ copies of the initial state. In the next subsection we shall introduce the general definition based on a more rigorous setup of quantum hypothesis testing. 

\subsection{Definition}\label{sec:def}

In this subsection we provide the rigorous reasoning behind our definition of STMI, and present the general definition. We begin by reviewing the hypothesis testing interpretation of relative entropy \cite{hiai1991proper,hayashi2001asymptotics}. To this aim, consider a black box that may contain either $N$ copies of quantum state $\rho$, or $N$ copies of quantum state $\sigma$. We are allowed to perform arbitrary measurements on this $N$-copied system to tell whether it is $\rho^{\otimes N}$ or $\sigma^{\otimes N}$. Now, if we make a hypothesis that the state is $\sigma^{\otimes N}$, and carry out some measurement, we can compute the probability of a measurement output {\it assuming} the state is $\sigma^{\otimes N}$. If our hypothesis is correct, this probability will reach $1$ in large $N$, but if our hypothesis is incorrect, {\it i.e.} if the state is actually $\rho$, then the typical result has a smaller probability $P_N\left(\rho|\sigma\right)$, which is lower bounded by the relative entropy: $P_N\left(\rho|\sigma\right)\geq e^{-NS\left(\rho|\sigma\right)}$. Here, $S(\rho|\sigma)={\rm tr}\left(\rho \log\rho-\rho\log\sigma\right)$. A smaller probability means that one can conclude the hypothesis is wrong with a higher confidence. Therefore, the relative entropy provides the fastest rate at which one can identify a wrong hypothesis. If $\rho$ and $\sigma$ are close to each other, $S(\rho|\sigma)$ is small and it will be difficult to distinguish them. In contrast, if $\sigma$ is not full rank, there are states which never appear in $\sigma$. For example, in a qubit system with states $|0\rangle,|1\rangle$, if $\sigma=|0\rangle\langle 0|$, then the probability of seeing $|1\rangle$ is zero. Thus if the probability of $|1\rangle$ is nonzero in $\rho$, if we perform a measurement in this basis we are certain that the state is $\rho$ the moment we observe $|1\rangle$, which means $P_N=0$ for a finite $N$. This corresponds to a diverging relative entropy.

\begin{figure}
    \centering
    \includegraphics[width=5.5in]{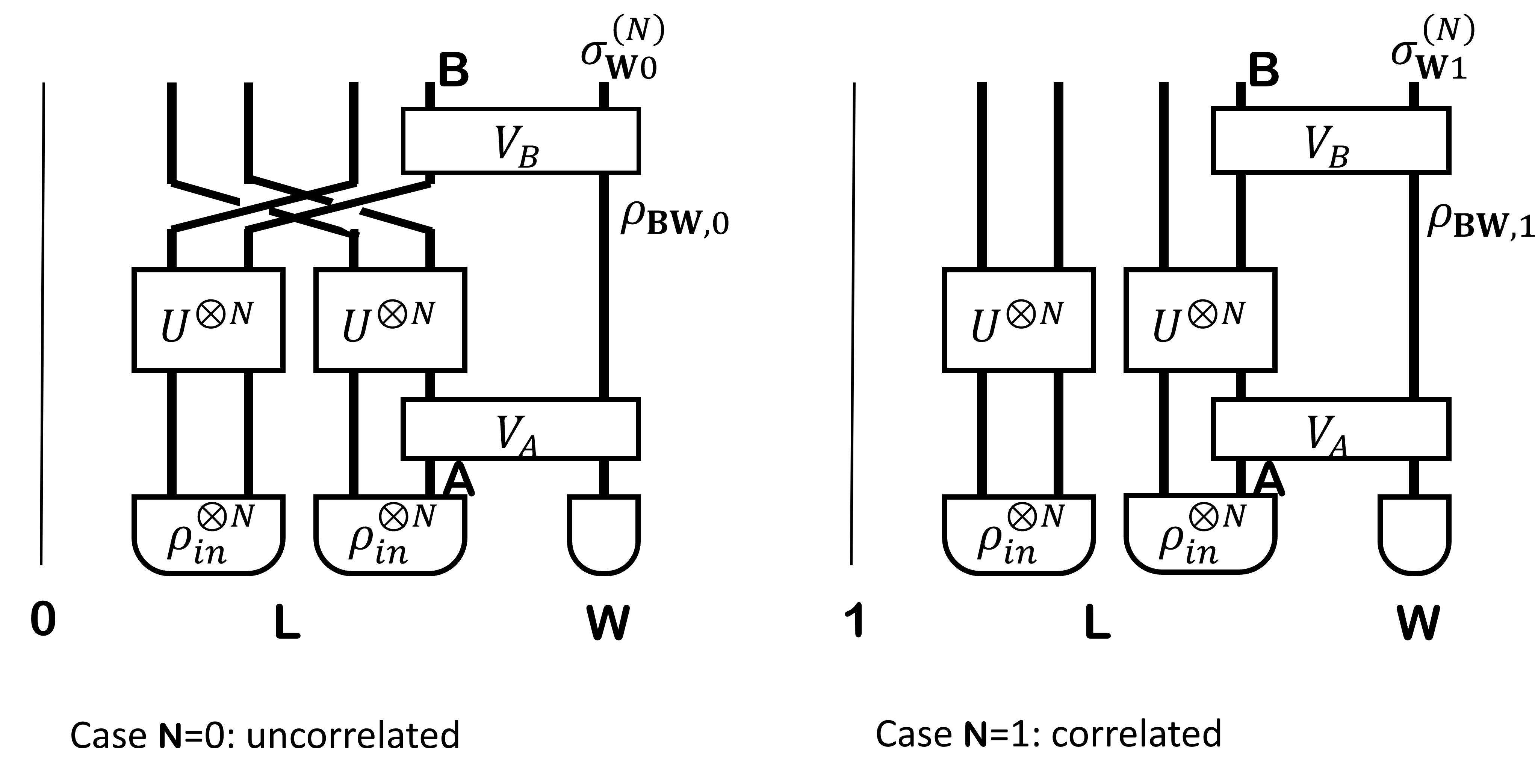}
    \caption{Definition of $J_{N}(\fA:\fB)$ in Eq. (\ref{eq:def1}) for the uncorrelated and correlated case. 
    }
    \label{fig:def1}
\end{figure}

For two spacelike separated regions $\fA,\fB$, the mutual information $I(\fA:\fB)$ is a relative entropy $ I(\fA:\fB)=S\left(\rho_{\fA\fB}|\rho_{\fA}\otimes\rho_{\fB}\right)$. Thus mutual information determines the probability that $\rho_{\fA\fB}$ is mistaken to be the uncorrelated state $\rho_{\fA}\otimes\rho_{\fB}$. Inspired by this intepretation, as a space-time generalization we consider the two situations illustrated in Fig. \ref{fig:def1}. In both cases, the experimentalist controls the ancilla $\fW$ and the gates $V_A,V_B$ which couple $\fW$ with the physical system $\fL$. This coupling characterizes the most general experiments that can occur, which includes measurements of correlation functions in the original system, and also includes more general quantum processes. For example, we can swap a qubit in $\fA$ with a qubit in $\fW$ and later measure the swapped-out qubit together with $\fB$ in an entangled basis. In situation $1$, $\fW$ is coupled with the same $N$ copies of $\fL$ at the space-time region $\fA$ and $\fB$. In situation $0$, the coupling at $\fB$ occurs with a new set of $N$ copies of systems, such that there won't be any correlation between $\fB$ and $\fW$. The experimentalist does not know whether it is situation $0$ or $1$. If they make a conjecture that it is situation $0$, and actually it is situation $1$, the rate of finding out that the conjecture is wrong is determined by the relative entropy of the output state of $\fW$, denoted as $\sigma_{\fW 1}^{(N)}$ and $\sigma_{\fW 0}^{(N)}$ respectively. Thus it is natural to define the space-time mutual information $J(\fA:\fB)$ by optimizing this relative entropy over the choice of $V_A,V_B$:

\begin{align}
J_N(\fA:\fB)=\frac1N\sup_{V_A,V_B}S\kc{\sigma_{\fontH{W},1}^{(N)}\middle|\sigma_{\fontH{W},0}^{(N)}}\label{eq:def1},\qquad J(\fA:\fB)=\lim_{N\to\infty}J_N(\fA:\fB)
\end{align}

More formally, this setup is an example of quantum algorithmic measurement (QUALM) defined in Ref. \cite{aharonov2021quantum}. The two situations in Fig. \ref{fig:def1} are denoted as two ``lab oracles,'' where the intrinsic dynamics $U$ is given by nature, while the experimentalist has the option of choosing the coupling $V_A,V_B$ between $\fL$ and $\fW$. A QUALM refers to an algorithm, {\it i.e.} a choice of gates $V_A,V_B$ for the purpose of achieving a particular task, similar to how a quantum algorithm is chosen to achieve a certain classical computation. In our case, the task is to distinguish the two lab oracles, with the optimal QUALM obtained by maximizing the relative entropy between the two output states.

The definition (\ref{eq:def1}) can be simplified by realizing that the supremum over $V_B$ can be achieved explicitly. Denote the state before applying $V_B$ for $\fB$ and $\fW$ as $\rho_{\fB^N\fW, a},~a=0,1$. One can see that $\sigma_{\fW, a}^{(N)}$ is related to $\rho_{\fB^N\fW, a}$ by a quantum channel induced by $V_B$ followed by a partial trace over $\fB$:
\begin{align}
    \sigma^{(N)}_{\fW, a}=\mathcal{C}\left(\rho_{\fB^N\fW, a}\right)
\end{align}
Due to the monotonicity of relative entropy under quantum channels, we have
\begin{align}
    S\left(\sigma_{\fW,1}^{(N)}|\sigma_{\fW,0}^{(N)}\right)\leq S\left(\rho_{\fB^N\fW,1}|\rho_{\fB^N\fW,0}\right)
\end{align}
More explicitly, we can define $\fW=\fW_{\fA}\fW_{\fB}$, and only $\fW_{\fA}$ is acted upon by the coupling $V_A$. $\fW_{\fB}$ has an initial state that is in direct product with $\fW_{\fA}$, and it has the same size as $\fB$. In this situation, $\rho_{\fB^N \fW, a}=\rho_{\fB^N \fW_{\fA}, a}\otimes \rho_{\fW_{\fB}}$. 
If we take $V_B$ to be a swap between $\fW_{\fB}$ and $\fB$, $\sigma_{\fW, a}^{(N)}$ is identical to $\rho_{\fB^N \fW_{\fA}, a}$, so that the relative entropy is the same. Therefore the swap operator achieves the optimization over $V_B$, and we can directly define the STMI using the state $\rho_{\fB^N\fW, a}$:
\begin{align}
    J_N(\fA:\fB)=\frac1N\sup_{V}S(\rho_{\fB^N\fW}|\rho_{\fB,0}^{\otimes N}\otimes \rho_{\fW})\qquad J(\fA:\fB)=\lim_{N\to\infty}J_N(\fA:\fB),\label{eq:def2}
\end{align}
where from now on we shall drop the subscript 1 from the connected state $\rho_{\fB^N\fW}$ for ease of notation, and we shall also drop the subscript $\fA$ from $W$ and $V$. In the above equation we have used the fact that $\rho_{\fB^N\fW, 0}$ is by design in a factorized form $\rho_{\fB^N\fW, 0}=\rho_{\fB, 0}^{\otimes N}\otimes \rho_{\fW}$, where $\rho_{\fW}={\rm tr}_{\fB^N}\left(\rho_{\fB^N\fW}\right)$ is determined by $\rho_{\text{in}}$ and $V$.

As a side remark, one special choice of the ancilla and its coupling to subsystem $\fA$ is to choose $\fontH{W}=\fontH{W}_1\fontH{W}_2$, in which $\fW_1$ and $\fW_2$ each has the same Hilbert space dimension as that of $A$, 
 and they are prepared in a maximally entangled state with each other. If we choose $V$ to be a SWAP gate between $\fontH{A}$ and $\fW_1$, $\rho_{\fB\fW}$ is equivalent to the superdensity operator defined in Ref. \cite{cotler2018superdensity}. 

\subsection{Properties of Space-time Mutual Information}\label{sec:properties}

We now explore a few basic properties of $J(\fA:\fB)$. 

{\bf Maximal size of $\fW$.} In our definition of STMI, there is no restriction on the size of ancilla $\fW$. However, $\fW$ does not need to be arbitrarily large. Without lost of generality, we can assume the initial state of $\fA\overline{\fA}$ and that of $\fW$ to be pure, $|\Phi_{\fA\bar{\fA}}\rangle$ and $|\Phi_{\fW}\rangle$. If that is not the case, we can always enlarge the system and introduce its purification. For simplicity of notation let us consider the case with one copy. After applying unitary $V$ we obtain a pure state $\ket{\Psi_{\fA\overline{\fA}\fW}}=V\ket{\Phi_{\fA\overline{\fA}}}\otimes\ket{\Phi_{\fW}}$. Expand this state in an arbitrary basis $\ket{a_{\fA}}$ of $\fA$, we can express
\begin{align}\label{eq:bound}
    \ket{\Psi_{\fA\overline{\fA}\fW}}=\sum_{a,b=1}^{d_\fA}\ket{b}_{\fA}\otimes\braket{a_{\fA}}{\Phi_{\fA\overline{\fA}}}\otimes V_{ba}\ket{\Phi_\fW}
\end{align}
This expression makes it explicit that the rank of the reduced state of $\fW$ is always bounded by $d_\fA^2$ with $d_\fA$ the Hilbert space dimension of $\fA$. Therefore it is always sufficient to take $d_\fW=d_\fA^2$. The discussion here easily generalizes to $N$ copies, in which case it is sufficient to take $d_\fW=d_\fA^{2N}$. In other words, the size of $\fW$ (number of qudits) can be taken as $2N$ copies of $\fA$. Additionally, by arbitrariness of $V$, we can choose $|\Phi_{\fW}\rangle$ to be the EPR state (or any other fixed reference state).


{\bf An alternative expression.} 
We can decompose $J(\fA:\fB)$ into two terms:
\begin{align}
    J_N(\fA:\fB)&=\sup_{V}\frac1N\kd{-S\kc{\rho_{\fontH{B}^N\fontH{W}}}-{\rm tr}\rho_{\fontH{B}^N\fontH{W}}\kc{\log \rho_{\fontH{B},0}^{\otimes N}+\log \rho_{\fontH{W}}}}\\
&=\sup_{V}\frac1N\kd{I^{(N)}\kc{\fontH{B}:\fontH{W}}+S\left(\rho_{\fontH{B}^N}\middle|\rho_{\fontH{B},0}^{\otimes N}\right)}\label{eq:def3}
\end{align}
The first term is the ordinary mutual information in state $\rho_{\fB^N\fW}$, while the second term is the relative entropy between the state of $\fB$ with and without the coupling with $\fW$. Here $\rho_{\fB}^{(N)}={\rm tr}_{\fW}\left(\rho_{\fB^N\fW}\right)$. The second term is a consequence of the causal influence of $\fA$ on $\fB$, which highlights the fact that even if we do not access the ancilla $\fW$, the coupling with $\fW$ still has nontrivial effect on the state of $\fB$. Physically, these two terms 
represent two ways to distinguish the correlated state $\rho_{\fontH{B}^N\fontH{W}}$ and the uncorrelated state $\rho_{\fB,0}^{\otimes N}\otimes \rho_{\fW}$. The second term is sensitive to how much the coupling can change the state of $\fB$, while the first term implies that even if the state of $\fB$ does not change, one can still measure the correlation between $\fA$ and $\fB$ by measuring that between $\fB$ and $\fW$. For example, let us consider a simple case when the evolution operator is trivial, and $\fA$ and $\fB$ are the same spatial region. If initially $\fA$ and $\overline{\fA}$ are in a maximally entangled EPR pair state, and $\fW$ is in a maximally entangled EPR pair state of ancilla subregions $\fW_1$ and $\fW_2$, each with the same dimension as $\fA$, then we can take $V$ to be a SWAP gate between $\fA$ and $\fW_1$, after which $\rho_{\fB}=\rho_{\fB,0}$, so that the second term vanishes, but the first term is nonzero. Alternatively, if we prepare $\fW_1$ in a pure state and still apply a SWAP gate, the second term will be nonzero, while the first term is smaller. In general, the maximization over $V$ is achieved by a compromise between these two terms.

{\bf Reduction to ordinary mutual information.} If $\fontH{B}$ and $\fontH{A}$ are space-like separated so that we can define a quantum state $\rho_{\fA\fB}$, gates applied to $\fontH{A}$ will never affect the state of $\fontH{B}$, so that $\rho_{\fontH{B}^N}=\rho_{\fontH{B},0}^{\otimes N}$. In this case the second term in (\ref{eq:def3}) vanishes. In addition, the mutual information $I^{(N)}\kc{\fontH{B}:\fontH{W}}$ satisfies the monotonicity $I^{(N)}\kc{\fontH{B}:\fontH{W}}\leq NI(\fontH{A}:\fontH{B})$, where $I(\fontH{A}:\fontH{B})$ is defined in the original system. Furthermore, the equal sign is achieved by choosing $V$ to be a swap gate, so that we obtain $J_N(\fA:\fB)=I(\fontH{A}:\fontH{B})$.

{\bf Monotonicity.} The mutual information is monotonously non-increasing when a quantum channel is applied to $\fA$ or $\fB$ separately. The same applies to STMI. For region $\fB$, this follows directly from the monotonicity of relative entropy \cite{nielsen2000quantum}. For a generic quantum channel $\mathcal{N}_\fB$ applied to $\fB$, we have
\begin{align}
    S\left(\mathcal{N}^{\otimes N}_\fB\kc{\rho_{\fB^N\fW}}\middle|\mathcal{N}_{\fB}\kc{\rho_{\fB,0}}^{\otimes N}\otimes\rho_\fW\right)\leq S\left(\rho_{\fB^N\fW}\middle|\rho_{\fB,0}^{\otimes N}\otimes\rho_\fW\right)
\end{align}
so that $J(\fA:\fB)$ is non-increasing.

For a quantum channel $\mathcal{N}_\fA$ applied to $\fA$, the proof is slightly more nontrivial. One can consider the dilation of $\mathcal{N}_\fA$, i.e. an isometry $K$ from $\fA$ to a bigger system $\fA\otimes \tilde{\fW}$. $\mathcal{N}_\fA$ is obtained by applying this isometry followed by tracing over $\tilde{\fW}$, {\it i.e.} $\mathcal{N}_\fA\left(\rho_\fA\right)={\rm tr}_{\tilde{\fW}}\left(K\rho_{\fA}K^\dagger\right)$. Computing $J_N(\fA:\fB)$ after applying the channel $\mathcal{N}_\fA$ requires to apply $V$ (which couples $\fA$ and $\fW$) after applying $K$. Then we can merge $\tilde{\fW}$ with $\fW$ and view $V\cdot K$ as a coupling between $\fA$ and a bigger ancilla $\tilde{\fW}\otimes\fW$. Therefore for arbitrary $V$, we can apply the monotonicity of relative entropy by tracing over $\tilde{\fW}$ to obtain
\begin{align}
    S\left(\rho_{\fB^N\fW\tilde{\fW}}\middle|\kc{\rho_{\fB,0}}^{\otimes N}\otimes\rho_{\fW\tilde{\fW}}\right)\geq S\left(\rho_{\fB^N\fW}\middle|\rho_{\fB,0}^{\otimes N}\otimes\rho_\fW\right)\,.
\end{align}
Taking the supremum over $V$ of the right-hand side yields
the STMI $J_N(\fA:\fB)$ after applying quantum channel $\mathcal{N}_\fA$, while the left-hand side, upon optimization over $V$, yields the STMI without applying $\mathcal{N}_\fA$. This implies monotonicity of $J_N(\fA:\fB)$ with respect to the application of a quantum channel in either $\fA$ or $\fB$. 

As a special case of the above, $J_N(\fA:\fB)$ is non-increasing upon tracing over part of $\fA$ or $\fB$. In other words, for disjoint regions $\fA,\fontH{C}$ at the initial time, and disjoint regions $\fB,\fontH{D}$ at the final time, we have $J(\fA:\fB)\leq J(\fA\fontH{C}:\fB)$, $J(\fA:\fB)\leq J(\fA:\fB\fontH{D})$. 

For the ordinary mutual information, when the inequality above is saturated, {\it i.e.} $I(\fA\fontH{C}:\fB)=I(\fA:\fB)$, the state $\rho_{ABC} $ satisfies the Markov property and can be reconstructed from its marginal on $\fA\fB$. We obtained similar results for the STMI, which we will discuss in Sec. \ref{sec:conditional STMI}.




{\bf Absence of upper bound.} In Eq. (\ref{eq:def3}), the mutual information term is always finite, upper bounded by $2\log d_{\fontH{B}}$. The relative entropy term could diverge. The divergence occurs if $\rho_{\fontH{B},0}$ is not full rank, and $\rho_{\fontH{B}^N}$ has a nonzero probability to be in the null space of $\rho_{\fontH{B},0}^{\otimes N}$. For example, if there is a conserved charge and the original state $\rho_{\fontH{B},0}$ is supported in a charge range $\kd{q_1,q_2}$, then as long as we can tune $V$ to change the charge of $\fontH{B}$ to be beyond this range, $J(\fA:\fB)$ diverges. If $\rho_{\fB,0}$ is full rank, with a nonzero minimal eigenvalue $p_{\rm min}$, one can prove that 
\begin{align}
    S\left(\rho_{\fB^N}\middle|\rho_{\fB,0}^{\otimes N}\right)\leq \log \frac1{p_{\rm min}}-S\left(\rho_{\fB^N}\right)\leq \log \frac1{p_{\rm min}}
\end{align}
Thus we obtain $J(\fA:\fB)\leq 2\log d_{\fB}+\log \frac1{p_{\rm min}}$.

\begin{figure}
    \centering
    \includegraphics[width=1.6in]{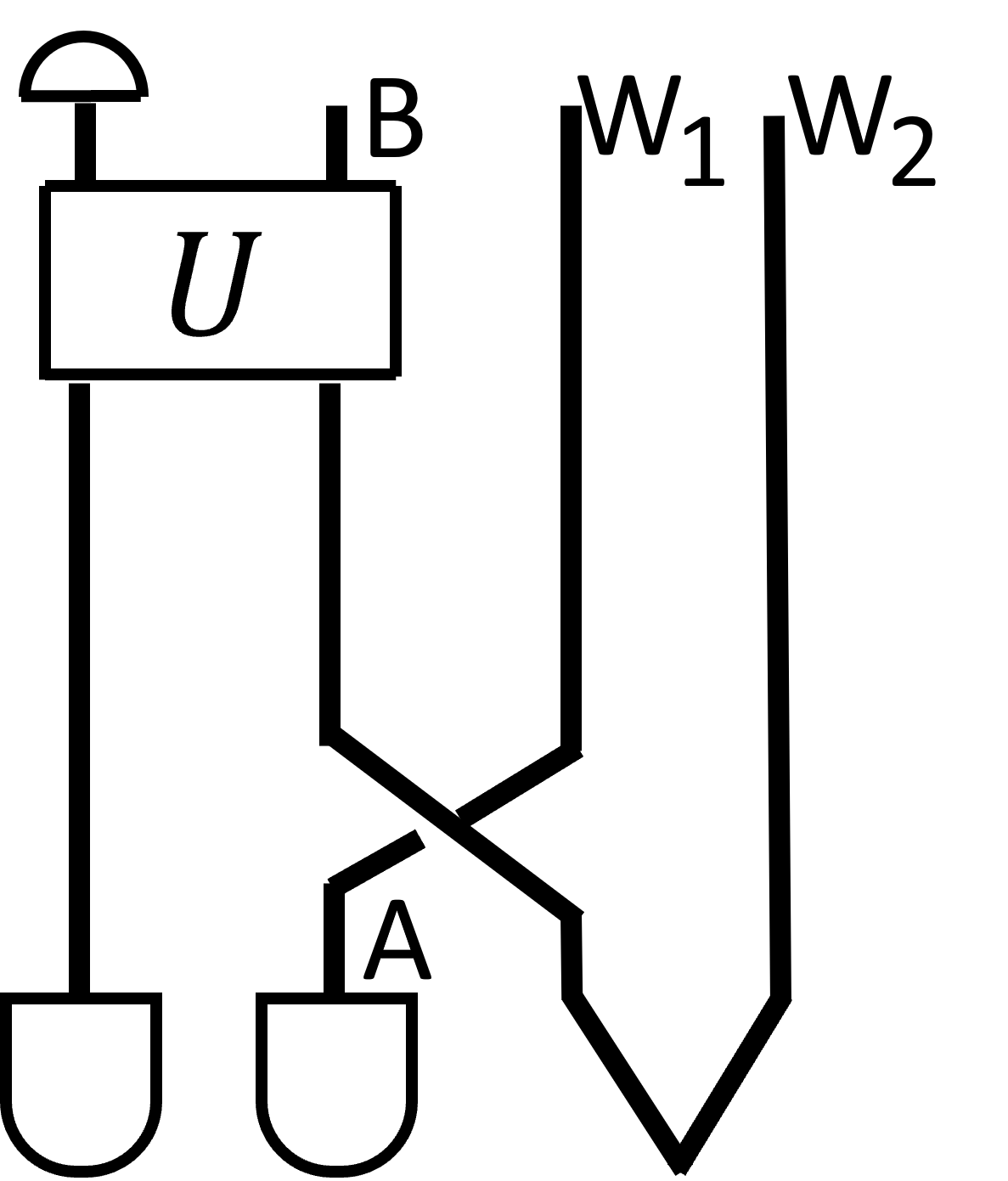}
    \caption{Mutual information of the Choi state corresponding to the unitary $U$.}
    \label{fig:choi}
\end{figure}

{\bf Relation to Choi state mutual information.} A related quantity to the STMI is the mutual information of the Choi state corresponding to a given unitary evolution which, in terms of Fig. \ref{fig:choi}, is given by $I(\fB:\fW_2)$ \cite{hosur2016chaos}. Such quantity can be viewed as the mutual information term in (\ref{eq:def3}) where we choose $V$ to be the SWAP between $\fA$ and $\fW_2$, with $\fW=\fW_1\fW_2$. We then infer that the Choi state mutual information cannot be larger than the STMI:
\be \label{choiin}I(\fB:\fW_1)\leq J(\fA:\fB)\,.\ee
Additionally, note that in Fig. \ref{fig:choi} $\bar {\fA}$ is in the infinite-temperature state. The space-time mutual information $J(\fA:\fB)$, on the other hand, is applicable for any initial state, including states where $\fA$ and $\bar{\fA}$ are entangled.

{\bf Vanishing condition.} The condition $J_N(\fA:\fB)=0$ requires $\rho_{\fontH{B}^N\fontH{W}}=\rho_{\fontH{B},0}^{\otimes N}\otimes \rho_{\fontH{W}}$ for all $V$. This obviously requires all correlation functions in $\fA$ and $\fB$ to be disconnected. Conversely, assume that all (Keldysh-ordered) correlation functions between $\fA$ and $\fB$ factorize, i.e. (taking $N=1$ for simplicity)
\be \tr\left(\mathcal O_{\fB}U\mathcal O_{\fA1}\rho_{\text{in}}\mathcal O_{\fA2}U^\dag\right)=\tr\left(\mathcal O_{\fB}U\mathcal \rho_{\text{in}}U^\dag\right)\tr\left(O_{\fA1}\rho_{\text{in}}\mathcal O_{\fA2}\right)\,.\ee
Then, the $\fB\fW$ correlation function in the system coupled with ancilla $W$ is
\be\begin{split}&\tr\left(\mathcal O_B \mathcal O_W U V\rho_{\text{in}}V^\dag U^\dag\right)=\sum_{ij}\tr\left(\mathcal O_B  U V_i\rho_{\text{in}}V_j^\dag U^\dag\right)\mathcal O^{ji}_W\\
&=\sum_{ij}\tr\left(\mathcal O_B U\rho_{\text{in}}U^\dag\right)
\tr\left(V_i \rho_{\text{in}}V_j^\dag\right)\mathcal O_W^{ij}
=\tr\left(\mathcal O_B U\rho_{\text{in}}U^\dag\right)\tr\rho_{\text{in}}\mathcal O_W
\end{split}\ee
also factorizes, where we decomposed the isometry $V=\sum_i V_i|i\rangle$, with $|i\rangle$ a basis on $\fW$ and $V_i$ an operator acting on $\fA$, and $\mathcal O^{ji}_W=\langle j |\mathcal O_W| i\rangle $. This in turn implies that $\rho_{\fB\fW}$ necessarily factorizes, $\rho_{BW}=\rho_B\otimes\rho_W$, with $\rho_B=\rho_{B,0}$, thus implying $J_1(\fA:\fB)=0$. Similarly, when we couple $N$ copies of $A$ to $W$, we can also express the $\fB\fW$ correlation as a sum over Keldysh-contour ordered correlators of $\fA$ and $\fB$, which in turn proves that $J_N(\fA:\fB)=0$ if all $\fA\fB$ correlation functions factorize.

{\bf Thermodynamics.} If $U$ is a chaotic time evolution $U=e^{-iHt}$ and $\fB$ is smaller than half of the system size, for long enough $t$ the subregion $\fB$ will reach thermal equilibrium, so that the only dependence of $\rho_{\fontH{B}}$ on $V$ will by given by thermodynamic quantities. For simplicity let us consider a system with energy conservation and no other conservation law. In this case, the only possible change caused by the coupling $V_A$ with the ancilla is the change of temperature. Thus we have $\rho_{\fB^N}=\rho^{\otimes N}_{\beta'}$ and $\rho_{\fB,0}^{\otimes N}=\rho^{\otimes N}_{\beta}$ are thermal states at temperatures $\beta'$ and $\beta$, respectively. Here we have assumed that $V_A$ does not cause a large energy fluctuation in the system, such that $\beta'$ is well-defined. In this case, the mutual information $I(B^N:W)$ is negligible, and the main contribution to STMI comes from the relative entropy term:
\begin{align}
J(A:B)\simeq S\kc{\rho_{\beta'}\middle|\rho_\beta}=\beta\kc{\avg{H}_{\beta'}-\avg{H}_\beta}-\kc{S_{\beta'}-S_\beta}=\beta\Delta F_\beta \label{asymptfin}
\end{align}
This is the change of thermal free energy $F_\beta[\rho]=\langle H\rangle_\rho-\beta^{-1}S(\rho)$. 

More generically, $V_A$ can cause a large energy fluctuation. For example, consider a single copy of the system coupled with $W$, and $V_A$ can create a Schroedinger cat state in the system, so that the reduced density matrix of $BW$ is
\begin{align}
    \rho_{BW}=\rho_{\beta'}\otimes \ket{0}\bra{0}+\rho_{\beta''}\otimes\ket{1}\bra{1}
\end{align}
where $0$ and $1$ are states of $W$. If this happens, there will be a nontrivial classical mutual information $I(B:W)$. Since energy is the only variable that $W$ can correlate with, and quantum correlation cannot be preserved in $B$ after a chaotic evolution, the mutual information is limited by the energy uncertainty. If $V_A$ has the ability of changing the energy of $B$ by $\Delta E$ at most, then the mutual information is upper-bounded by 
\begin{align}
I(B:W)\lesssim \log\frac{\Delta E}{\delta E}
\end{align}
with $\delta E=\sqrt{\avg{H^2}_{\fontH{B}}-\avg{H}^2_{\fontH{B}}}$
the intrinsic energy uncertainty. We expect this to be a small contribution. If we consider the limit that $B$ is a finite (smaller than half) portion of the entire system, and consider the thermodynamic limit, then at most the mutual information term will be proportional to $\log|B|$, while the relative entropy term is proportional to $|B|$ as long as $\beta'\neq \beta$.


\section{Bound on space-time correlation functions}\label{sec:bound}
An important property satisfied by the STMI is that, as we will now show, it bounds any two-point function between two possibly causally connected subregions. This generalizes the bound of the standard mutual information of spatial correlation functions \cite{wolf2008area}.
\begin{theorem}\label{thm}
The STMI bounds all two-point correlation functions between subsystems $\fA$ and $\fB$. Explicitly, for any $N\geq 1$, we have the following bounds on symmetric and retarded correlation functions:
\begin{align}\label{eq:ret}
J_N(\fA:\fB)&\geq \frac 18\left(\frac{-i\tr\rho_{\text{in}}[\mathcal O_{\fB}(t),\mathcal O_{\fA}]}{||\mathcal O_{\fA}||_{\infty}||\mathcal O_{\fB}||_{\infty}}\right)^2\\
J_N(\fA:\fB)&\geq\frac 18\left(\frac{\left(\tr\rho_{\text{in}}\{\mathcal O_{\fB}(t),\mathcal O_{\fA}\}\right)_c}{||\mathcal O_{\fA}||_{\infty}||\mathcal O_{\fB}||_{\infty}}\right)^2\, \label{eq:sym}
\end{align}
where $\mathcal O_{\fB}(t)$ is a (Heisenberg) operator supported in subregion ${\fB}$, and similarly for $\mathcal O_{\fA}$, $||\cdot||_{\infty}$ denotes the operator norm, $(\cdots)_c$ the connected component of a correlator, and $\mathcal O_{\fA},\mathcal O_{\fB}$ are assumed to be Hermitian. 
\end{theorem}
The numerators on the right-hand sides of the above inequalities correspond to retarded and connected symmetric two-point functions; these two correlators are sufficient to linearly generate any other Keldysh time ordering (such as Feynman, time-ordered correlators, etc.).

To prove the theorem, it is sufficient to consider the single-copy STMI $J_1(\fA:\fB)$. We take a specific choice of ancilla-system coupling, defined by
\be\label{Xw} \rho_{\text{in}}\otimes(|0\rangle\langle 0|)_{\fW}
\to \sum_{ij} X_i\rho_{\text{in}}X_j^\dag\otimes(|i\rangle\langle j|)_{\fW}\,,
\ee
where $i,j=0,1,2$, and the operators $X_i$ act on $\fA$ and are defined as
\be X_0=\sqrt{\tfrac12}\mathbb1,\qquad X_1=\sqrt{\tfrac12} \frac{\mathcal O_{\fA}}{||\mathcal O_{\fA}||_{\infty}},\qquad X_2=\sqrt{\tfrac12}\sqrt{\mathbb1-\frac{\mathcal O_{\fA}^2}{||\mathcal O_{\fA}||^2_{\infty}}}\,.
\ee
These operators satisfy $\sum_i X_i^\dag X_i=\mathbb1$, thus guaranteeing that the coupling to the ancilla is an isometry, and can thus be extended to a unitary operator. The coupling (\ref{Xw}) can be thought of as a control-$\mathcal O_{\fA}$ gate and was suitably chosen so that, as we will see, it will reproduce the two-point function we want to bound. Now, to prove the bound for the retarded two-point function (\ref{eq:ret}), define the following operator acting on $\fW$:
\be Y_{\fW}=\begin{pmatrix}0&-i&0\\i&0&0\\0&0&0\end{pmatrix}\,.\ee
We then find that the retarded two-point function can be viewed as the following expectation value over the state $\rho_{\fB\fW}$:
\be \label{eq:traceY}\tr(\rho_{\fB\fW}Y_{\fW}\mathcal O_{\fB})=\sum_{ij}\tr\left[ U X_i\rho_{\text{in}} X_j U^\dag\otimes (|i\rangle\langle j|)_{\fW} Y_{\fW} \mathcal O_{\fB}\right]=-\frac i2\tr\rho_{\text{in}}[\mathcal O_{\fB}(t),\mathcal O_{\fA}]\,.\ee
Now consider the following sequence of inequalities
\be S(\rho_{\fB\fW}|\rho_{\fB 0}\otimes\rho_{\fW})\geq \frac 12||\rho_{\fB\fW}-\rho_{\fB 0}\otimes\rho_{\fW}||_1^2\geq \frac{1}{2}\left|\frac{\tr(\rho_{\fB\fW}-\rho_{\fB 0}\otimes\rho_{\fW})Y_{\fW}\mathcal O_{\fB}}{||Y_{\fW}||_{\infty}||\mathcal O_{\fB}||_{\infty}}\right|^2\,,\label{eq:ineq1}\ee
where in the first step we applied the quantum Pinsker's inequality, and in the second step we used H\"older's inequality. Comparing (\ref{eq:ineq1}) with (\ref{eq:traceY}), we arrive at (\ref{eq:ret}). Note that the disconnected state does not contribute to the trace on the right-hand side of (\ref{eq:ineq1}) due to the form of $Y_{\fW}$, 
which is equivalent to saying that the retarded two-point function does not have a disconnected component. The vanishing of this trace also implies that only the mutual information term in (\ref{eq:def3}) contributes to the bound, i.e. we have a tighter bound given by
\begin{align}\label{miboun1}
I(\fB:\fW)&\geq \frac 18\left(\frac{-i\tr\rho_{\text{in}}[\mathcal O_{\fB}(t),\mathcal O_{\fA}]}{||\mathcal O_{\fA}||_{\infty}||\mathcal O_{\fB}||_{\infty}}\right)^2\,.
\end{align}
Similar steps lead to the bound on the symmetric two-point function (\ref{eq:sym}), this time using, instead of $Y_{\fW}$,
\be\label{eqXW} X_{\fW}=\begin{pmatrix}0&1&0\\1&0&0\\0&0&0\end{pmatrix}\,.\ee
When $\fA$ and $\fB$ are causally disconnected, the symmetric two-point function reduces to the spatial one $\frac 12(\tr\rho_{\text{in}}\{\mathcal O_{\fB}(t),\mathcal O_{\fA}\})_c\to(\tr\rho_{\text{in}}\mathcal O_{\fB}\mathcal O_{\fA})_c$ and we recover the standard bound of \cite{wolf2008area}, with the correct numerical prefactor.

We emphasize that taking the supremum over $V$ in the definition of $J_1$ is crucial for the above proof of the bounds (\ref{eq:ret}),(\ref{eq:sym}). Indeed, consider e.g. fixing $V=S_{\fA \fW_1}\otimes\mathbb 1_{\fW_2}$, where $S_{\fA\fW_1}$ is the swap between $\fA$ and $\fW_1$. With this choice, $\rho_{\fB\fW}$ constitutes an example of superdensity operator defined in \cite{cotler2018superdensity}. In this case, as we show in appendix \ref{app:bound}, using similar steps as above, one can prove weaker bounds than (\ref{eq:ret}),(\ref{eq:sym}), with an overall dimensional suppression factor.


\section{Markov property}\label{sec:conditional STMI}

In the case of ordinary mutual information, for three regions $ABC$ one can define conditional mutual information
\begin{align}
    I(A:C|B)=I(A:BC)-I(A:B)
\end{align}
which is non-negative due to monotonicity of relative entropy under the action of quantum channels. $I(A:C|B)$ is the decrease of relative entropy $S(\rho_{ABC}|\rho_A\otimes\rho_{BC})$ under the partial trace over $C$. If $I(A:C|B)=0$, the state $\rho_{ABC}$ can be recovered from $\rho_{AB}$ and $\rho_{BC}$ using the Petz map \cite{petz_2003,hayden2004structure}:
\be\label{petzabc} \rho_{\fA\fB\fC}=\hat T\rho_{\fA\fB}\equiv\rho_{\fB\fC}^\frac 12\left(\rho_{\fB}^{-\frac 12}\rho_{\fA\fB}\rho_{\fB}^{-\frac 12}\otimes \text{Id}_{\fC}\right)\rho_{\fB\fC}^{\frac 12}\,.\ee
Since $\rho_{ABC}$ is determined by $\rho_{AB}$ and $\rho_{BC}$, correlation functions between $A$ and $BC$ can be determined by that between $A$ and $B$. More explicitly, for any operator $O_A$ supported on $A$ and $O_{BC}$ supported on $BC$, we can define
\begin{align}
    \tilde{O}_B=\rho_B^{-\frac12}\rho_{BC}^{\frac12}O_{BC}\rho_{BC}^{\frac12}\rho_B^{-\frac12}
\end{align}
such that 
\begin{align}
    {\rm tr}\left(\rho_{ABC} O_AO_{BC}\right)={\rm tr}\left(\rho_{AB}O_A\tilde{O}_B\right)\,.
\end{align}

\begin{figure}
    \centering
    \includegraphics[width=3in]{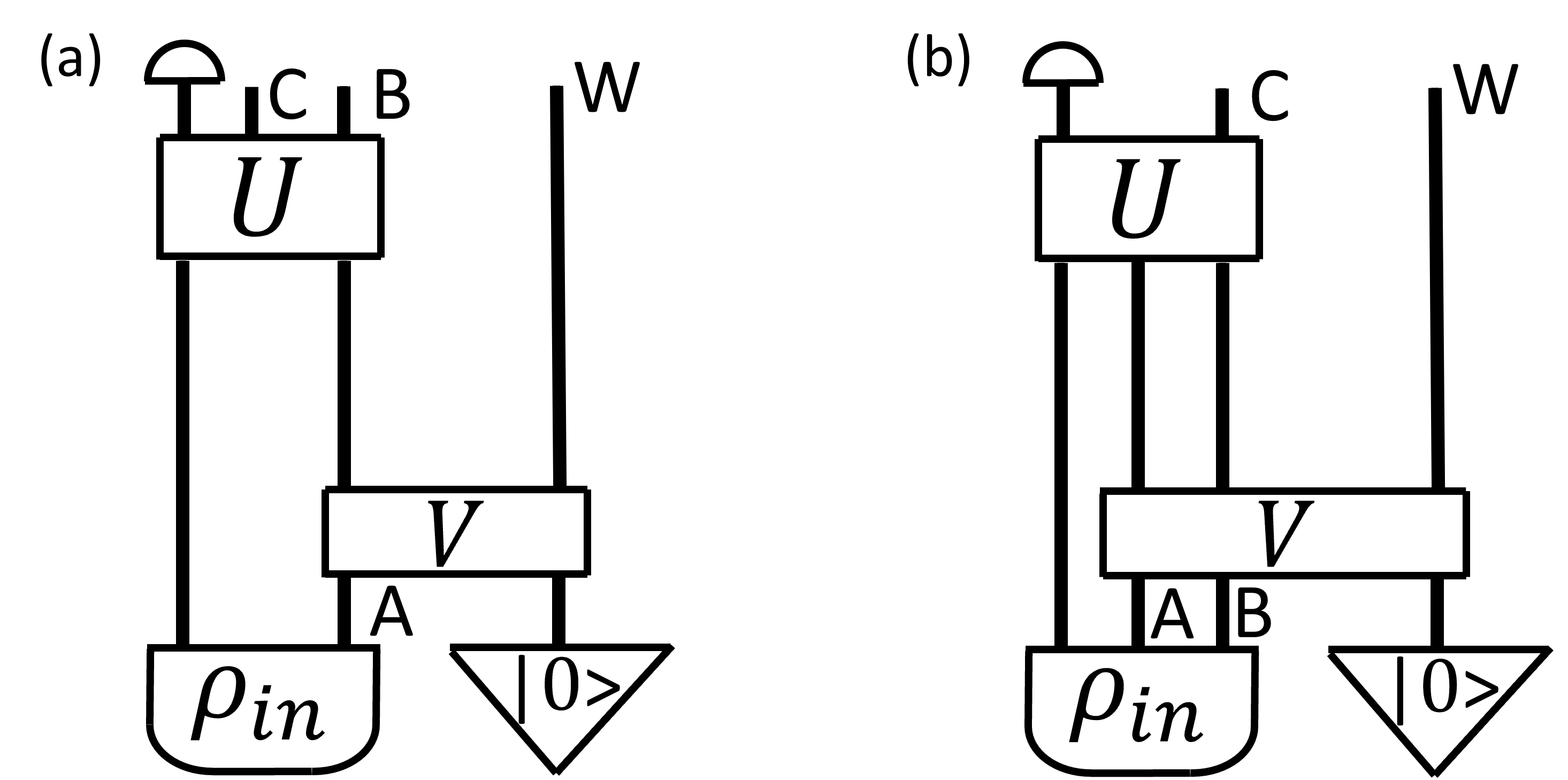}
    \caption{Illustration of the two situations involving three regions $ABC$, for the discussion of Markovian condition in Sec. \ref{sec:conditional STMI}.}\label{fig:markovian}
\end{figure} 
Now we consider the situation with spacetime mutual information. Consider three regions $A,B,C$ with $B$ and $C$ defined at the same future time, and $A$ defined at an earlier time, as is shown in Fig. \ref{fig:markovian} (a). Assume that 
\be 
\label{condJ} J_N(\fA:\fB\fC)=J_N(\fA:\fB)\,,
\ee
Naively, the unitary $V_1$ coupling $\fA$ and $\fW$ that optimize $J_N(\fA:\fB)$ may be different from the unitary $V_2$ that optimize $J_N(\fA:\fB\fC)$. However, they must actually be the same, which can be proven by contradiction. If $V_1\neq V_2$, we can take $V=V_1$ and compute the relative entropy 
\begin{align}
\tilde{J}_N(\fA:\fB\fC)\equiv \frac1N \left.S\left(\rho_{\fB^N\fC^N\fW}|\rho_{\fB\fC,0}^{\otimes N}\otimes\rho_W\right)\right|_{V=V_1}
\end{align}
If $V_1$ does not maximize this quantity, then we have 
\begin{align}\tilde{J}_N(\fA:\fB\fC)<{J}_N(\fA:\fB\fC)=J_N(A:B)\label{eq:JN_contradiction}
\end{align} 
Since $J_N(A:B)$ and 
$\tilde{J}_N(\fA:\fB\fC)$ are computed for the same system (with the same gate $V_1$), Eq. \eqref{eq:JN_contradiction} contradicts with monotonicity of relative entropy under partial trace over $C$. Therefore we have proven that for the same $V=V_1=V_2$, the relative entropies 
\begin{align}
S\left(\rho_{\fB^N\fC^N\fW}|\rho_{\fB\fC,0}^{\otimes N}\otimes\rho_W\right)=S\left(\rho_{\fB^N\fW}|\rho_{\fB,0}^{\otimes N}\otimes\rho_W\right)\,,
\end{align}
where both reach the maximum. Consequently, we can apply the Petz map and express
\be\label{petzbcw} \rho_{\fB\fC\fW}=\hat T\rho_{\fB\fW}\equiv\rho_{\fB\fC0}^\frac 12\left(\text{Id}_{\fC}\otimes \rho_{\fB0}^{-\frac 12}\rho_{\fB\fW}\rho_{\fB0}^{-\frac 12}\right)\rho_{\fB\fC0}^{\frac 12}\,.\ee
Interestingly, the map acts trivially on $W$. For any operator $O_{Bc}$ and $O_W$, there is a corresponding operator $\tilde{O}_B=\rho_{B0}^{-\frac12}\rho_{BC0}^{\frac12}O_{BC}\rho_{BC0}^{\frac12}\rho_{B0}^{-\frac12}$ such that ${\rm tr}\left(\rho_{BCW}O_{BC}O_W\right)={\rm tr}\left(\rho_{BW}\tilde{O}_BO_W\right)$. This inturn implies that any correlation function between $A$ and $BC$ that one can measure indirectly through measuring correlation between $BC$ and $W$ can actually be converted into a measurement that only involves $A$ and $B$. In other words, $C$ does not directly correlate with $A$. The correlation between $A$ and $C$ are only generated through $BC$ correlation and $AB$ correlation. This is exactly in parallel with the Markovian condition in the ordinary spatial mutual information case.

The other situation is shown in Fig. \ref{fig:markovian} (b), when regions $A$ and $B$ are at equal time and $C$ is at a later time. In that case, if we have $J_N(\fA\fB:\fC)=J_N(\fB:\fC)$, it simply means that the optimal coupling between $AB$ and $W$ that maximizes the relative entropy does not involve $A$. it is sufficient to couple the ancilla to $\fB$ in our optimization in order to find $J_N(\fA\fB:\fC)$, and $\fA$ remains untouched. In this case, $A$ is not a system that is traced out, and we cannot directly apply Petz map. A natural question is whether this condition also implies a Markov property on the joint ancilla-system state. We leave the exploration of this problem for future work.

\section{Quantum channel discrimination and additivity}
\label{sec:qcd}
\begin{figure}
    \centering
    \includegraphics[width=5.5in]{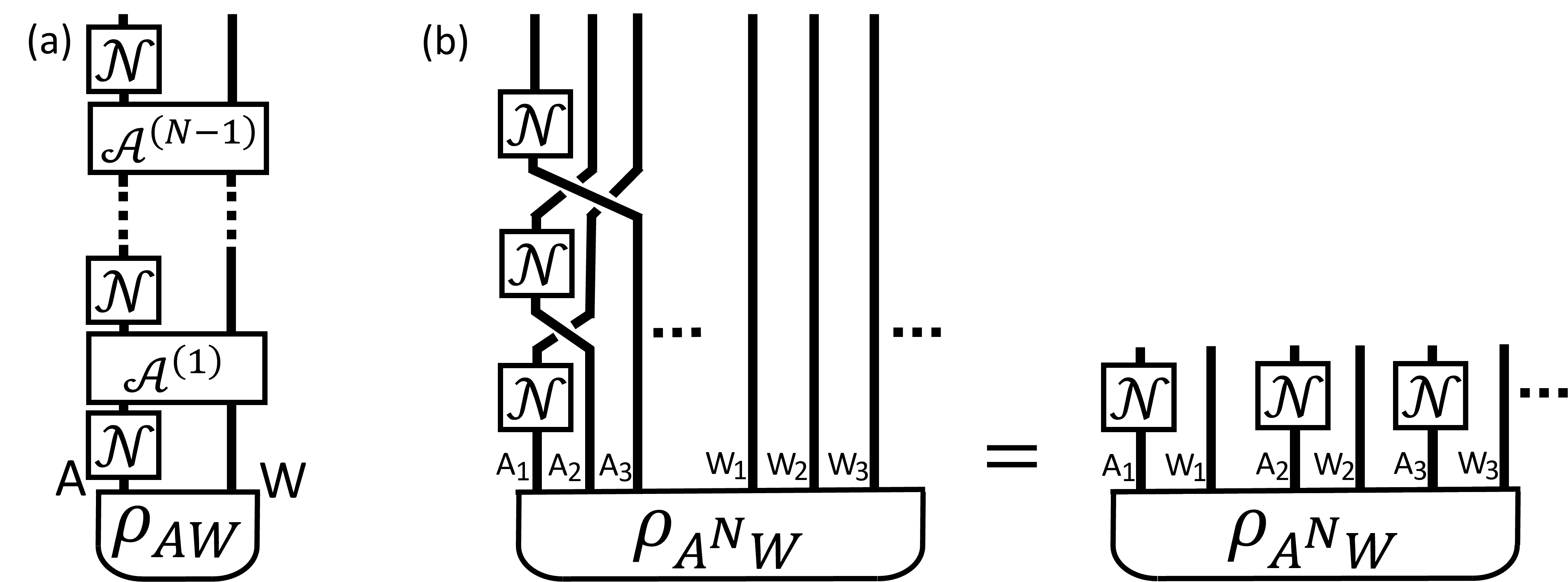}
    \caption{(a) Definition of $\mathcal N_N(\rhoAW)$. (b) By specifying the $\mathcal A^{(i)}$ to suitable swaps, $\mathcal N_N(\rhoAW)$ reduces to our (\ref{eq:def2}).}
    \label{fig:nnjn}
\end{figure}

Quantum channel discrimination can be viewed as a natural extension of quantum hypothesis testing, where the aim is to discriminate channels instead of states.  The formulation of this problem resembles that presented in Sec. \ref{sec:defi}, although, as we now illustrate, it is applied to a slightly different context. Consider two quantum channels $\mathcal N_1$ and $\mathcal N_2$, both with input and output systems $\fA$ and $\fB$, respectively. To discriminate whether the system measured has evolved through $\mathcal N_1$ or $\mathcal N_2$, one implements adaptive strategies consisting of alternating applications of the channel to discriminate and auxiliary channels. More explicitly, the application of $\mathcal N_1$ or $\mathcal N_2$ is alternated with channels $\mathcal A^{(i)}$ that map the output $\fB$, jointly with an ancilla $\fW$, to the input $\fA$ of the subsequent application of the channel, together with the ancilla, as shown in Fig. \ref{fig:nnjn}(a). After $N$ repetitions of this process, the final state is measured. From the formulation of the quantum hypothesis testing reviewed in Sec. \ref{sec:def} we know that the probability of incorrectly concluding that the channel is $\mathcal N_2$ is $e^{-NS(\rho^{(\mathcal A)}_1|\rho^{(\mathcal A)}_2)}$, where $\rho^{(\mathcal A)}_1$ and $\rho^{(\mathcal A)}_2$ are the output states that use channel $\mathcal N_1$ and $\mathcal N_2$, respectively, and the superscript $(\mathcal A)$ indicates that we  applied a given sequence of system-ancilla channels $\mathcal A^{(1)},\mathcal A^{(2)},\dots$, corresponding to a particular strategy. The best strategy is then characterized by the optimization
\be\label{qcd1} \sup_{\rho,\mathcal A^{(i)}}S(\rho^{(\mathcal A)}_1|\rho^{(\mathcal A)}_2)\,,\ee
where $\rho$ is the input state in Fig. \ref{fig:nnjn}(a). Due to the joint convexity of the relative entropy and to the fact that $\rho^{(\mathcal A)}_1$ and $\rho^{(\mathcal A)}_2$ are linear in $\rho$, it is sufficient to restrict to pure states $\rho$.

Quantum channel discrimination has many applications, e.g.: quantum illumination, to enhance the detection of targets in the presence of thermal noise through entangled photon pairs \cite{lloyd2008enhanced}; quantum metrology, to estimate unknown parameters of quantum channels \cite{pirandola2019fundamental}, and quantum reading, which involves the use of nonclassical transmitters to read data from classical digital memories \cite{pirandola2011quantum}.


\subsection{Additivity of the STMI for initial pure states}\label{sec:add1}

Let us now come back to the setup of Sec. \ref{sec:def}. Assuming the initial state reduced on $\fA$ is pure, i.e. $\rho_{\text{in}}=\rho_{\bar{\fA}}\otimes\rho_{\fA}$, with $\rho_{\fA}$ pure, the optimization problem (\ref{eq:def2}) reduces to a special case of quantum channel discrimination. To see the connection, note that both the system $\fA^{N}$ and the ancilla $\fW$ in eq. (\ref{eq:def2}) are in a pure state, and optimizing over $V$ corresponds to optimizing over the most general pure state of the joint system $\fA^N\fW$. One then writes the connected state $\rho_{\fB^N\fW}$ as $N$ tensor copies of channel $\mathcal N$ acting on the state $\rho_{A^NW}$ on subregion $\fA^N$, where channel $\mathcal N$ is obtained by tracing the time evolution in (\ref{eq:def2}) over $\bar {\fB}$. As shown in Fig. \ref{fig:nnjn}(b) this setup is then equivalent, for an appropriate choice of $\mathcal A^i$, to the setup of Fig. \ref{fig:nnjn}(a), where $\mathcal A^{(i)}$ simply swaps the $i$-th copy of $\fA$. For the disconnected state $\rho_{\fB,0}^{\otimes N}\otimes \rho_{\fW}$ we apply a similar reasoning, but instead of $\mathcal N$ we now have the replacer channel: $\mathcal R(\rho)=\rho_{\fB, 0}$ for any state $\rho$ of $\fA$. We then reduced (\ref{eq:def2}) to an instance of quantum channel discrimination, and can write
\be \label{jnqcd} J_N(A:B)=\frac 1N \sup_{\rho_{A^NW}} S\left(\mathcal N^{\otimes N}(\rho_{A^NW})|\mathcal R^{\otimes N}(\rho_{A^NW})\right)\ee
In particular, $J_1(\fA:\fB)$ identifies with the \emph{channel relative entropy} between $\mathcal N$ and $\mathcal R$ \cite{cooney-2016}. We will see in Sec. \ref{sec:constr} how, in the general case, the STMI \ref{eq:def2} can be formulated as a ``constrained'' quantum channel discrimination.

We now prove that the STMI $J_N(\fA:\fB)$, when $\rho_{\fA}$ is factorized, is independent of $N$, i.e. it is additive. Additivity is a fundamental question in quantum channel discrimination and, fortunately, this property was recently proven when the alternative hypothesis (i.e., the second argument in the relative entropy) is a replacer channel  \cite{cooney-2016}, which precisely corresponds to our setup. Theorem 1 of \cite{cooney-2016} essentially states that, for $N\to\infty$, (\ref{qcd1}) is equal to the channel relative entropy $\sup_{\rho_{\fA\fW}} S(\mathcal N_1\rho_{\fA\fW}|\mathcal N_2\rho_{\fA\fW})$ when $\mathcal N_2$ is a replacer channel. Using the setup of the previous paragraph with $\mathcal N_1=\mathcal N$ and $\mathcal N_2=\mathcal R$, we then have
\be \sup_{\rho_{AW}}S\left(\mathcal N\rho_{AW}|\mathcal R \rho_{AW}\right)
=\lim_N\frac 1N \sup_{\rho,\mathcal A}S\left((\mathcal N \rho_{AW})^{\mathcal A}|(\mathcal R\rho_{AW})^{\mathcal A}\right)
\geq J_N(A:B)\ee
where the equality comes from the theorem mentioned above, and the inequality comes from that $J_N(A:B)$ corresponds to a special choice of $\mathcal A$, as shown in \ref{fig:nnjn}(b). Moreover, since $\sup_{\rho_{AW}}S(\mathcal N\rho_{AW}|\mathcal R\rho_{AW})=J_1(A:B)$ due to (\ref{jnqcd}) and $J_N\geq J_1$, we conclude additivity $J_N=J_1$.

We observe that, in quantum channel discrimination, there exist counterexamples to additivity when the alternative hypothesis is a more general channel \cite{fang2020chain}.

We currently do not know if additivity holds for general initial states. In Appendix \ref{app:addgeneral} we show that the $N$-replicated optimization problem (\ref{eq:def2}) admits a stationary point (i.e., $V$ satisfies  $\delta_V S(\rho_{\fB^N\fW }|\rho_{\fB^N\fW, 0})=0$) of the form $V_N=V_1^{\otimes N}$, where $V_1$ is a stationary point for $J_1$. This is a necessary but not sufficient condition for additivity. 

\subsection{STMI as a constrained quantum channel discrimination}\label{sec:constr}
In the previous section we showed that when the initial state $\rho_A$ is pure, the STMI can be viewed as a quantum channel discrimination. In this section, we will discuss the situation when $A$ and $\bar A$ are entangled and in particular $\rho_A$ is not pure. We shall see that it is most natural to think of this case as a constrained quantum channel discrimination. First, we can assume that the initial state of the system $\rho_{\text{in}}$ is pure by extending $\bar{\fA}$ and tracing over the extension. We can then rewrite the STMI as
\be\begin{gathered}\label{Jt}  J_N(\fA:\fB)=\frac 1N \sup_{\rho\in \mathcal S}S(\mathcal N^{\otimes N}(\rho)|\mathcal R^{\otimes N}(\rho))\ ,\qquad \mathcal S=\{\rho:\tr_{\fonth A^N \fonth W}\rho=\rho_{\bar{\fA}}^{\otimes N}\}\ ,\end{gathered}\ee
where $\rho_{\bar{\fA}}=\tr_{\fonth A}\rho_{\text{in}}$, and $\rho$ is a state on $\fonth A^N{\bar{\fA}}^N\fonth W$. Also, $\mathcal N^{\otimes N}$ denotes $N$ tensor copies of channel $\mathcal N:\bar {\fA}\fA\to \fB$, and $\mathcal R$ is the replacer channel introduced in Sec. \ref{sec:add1}. Due to the joint convexity of the relative entropy and to the convexity of $\mathcal S$, it suffices to optimize over pure states.

One might ask if the constraint $\mathcal S$ in (\ref{Jt}) can be implemented as a quantum channel in such a way that (\ref{Jt}) can still be viewed as an unconstrained channel discrimination. More explicitly, we ask if there is a quantum channel $\mathcal Q:\fonth {DW}\to \bar{\fonth A}^{N}\fonth A^{N}\fonth W$, for some subsystem $\fonth D$, whose image $\mathcal Q(\fonth{DW})$ is exactly
$\mathcal S$ and acts trivially on $\fonth W$. The latter condition is necessary to maintain the structure of channel discrimination, i.e. $\fW$ plays the role of an idler on which no channel is applied. The answer to this question is negative. Indeed, any state of $\mathcal S$ can be written as $\rho=V\rho_{\text{in}}V^\dag$, where $V:\bar{\fA}\fonth{A}\to\bar{\fA}\fonth{A}\fW$ is an isometry acting trivially on $\bar{\fonth A}$, $\rho_{\text{in}}$ is a pure state, and for simplicity we are assuming $N=1$. This means that $\tr_{\bar{\fonth A}}\rho$ has a
fixed entanglement spectrum, independent of $V$. Assume that the quantum channel $\mathcal Q$ exists. Then, considering two arbitrary initial states $\sigma_1$ and $\sigma_2$ on
$\fonth{DW}$, the states $\rho_1=\tr_{\bar{\fA}}\mathcal Q(\sigma_1)$ and $\rho_2=\tr_{\bar{\fA}}\mathcal Q(\sigma_1)$ should have the same entanglement spectrum. If
their eigenstates are different, then $p\rho_1+(1-p)\rho_2=\tr_{\bar{\fonth A}}\mathcal Q(p\sigma_1+(1-p)\sigma_2)$ will not have the same eigenvalues for $0<p<1$, and we thus infer that $\rho_1=\rho_2$. This implies that $\tr_{\fonth D}\sigma_1 = \tr_{\fA}\rho_1=\tr_{\fA}\rho_2=\tr_{\fonth D}\sigma_2$, which contradicts the assumption that $\sigma_1$ and $\sigma_2$ are arbitrary states.

\section{An ansatz for factorized initial states}\label{sec:ansatz}

\begin{figure}
    \centering
    \includegraphics[width=5in]{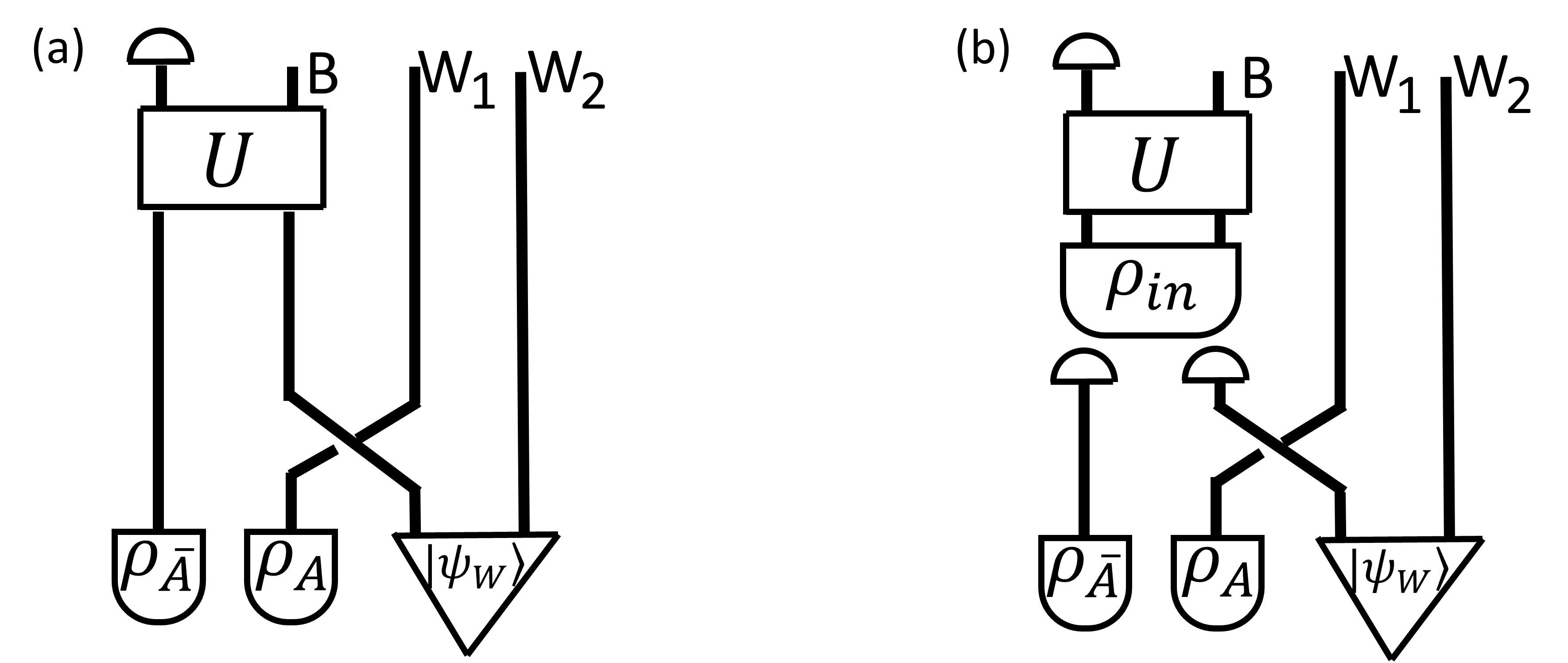}
    \caption{Representation of the the ansatz (\ref{eq:ansatz}) for the $N=1$ correlated (a) and uncorrelated (b) states $\rho_{BW}$ and $\rho_{B,0}\otimes \rho_W$.}
    \label{fig:ansatz}
\end{figure}

The optimization problem (\ref{eq:def2}) can be very non-trivial, especially for large Hilbert space dimension. In this Section we propose an ansatz leading to a simplification that applies when the initial state is factorized, i.e.
\be \label{eq:fact}\rho_{\text{in}}=\rho_{\bar{\fA}}\otimes\rho_{\fA}\,.\ee
We shall restrict to a single replica $N=1$. Keeping into account the bound $|\fW|\leq 2|\fA|$ from the discussion around eq. (\ref{eq:bound}), it is sufficient to take $\fW=\fW_1\otimes \fW_2$, with both $\fW_1$ and $\fW_2$ isomorphic to $\fA$. The ansatz consists of replacing $V$ with a swap $S$ between $\fA$ and $\fW_1$, and optimize over a generic initial state $|\psi_{\fW}\rangle$ of $\fW$:
\be J_N(\fA:\fB)=\frac 1N\sup_{\psi_W} S(\rho_{\fB^N\fW}|\rho_{\fB,0}^{\otimes N}\otimes \rho_{\fW})\,,\label{eq:ansatz}
\ee
as in Fig. \ref{fig:ansatz}.

We shall see below that the optimization reduces to a self-consistent equation for the reduced state of the ancilla $\fW_2$. This reduces the number of parameters to optimize from $2d_A^4-d_A^2$ (an isometry going from $\fA\to \fA\fW_1\fW_2$) to $d_{\fA}^2$ (number of independent mixed states on $\fW_2$). In Sec. \ref{sec:examples} we show several numerical examples leveraging this ansatz, including MBL and thermalizing systems.

As a support of this ansatz note that, when $\rho_{\fA}$ is pure, 
following a similar reasoning as that around (\ref{eq:bound}), we find that $|\fW|\leq |\fA|$. After applying the ancilla-system coupling $V$, we then have a pure state jointly defined on $\fA\fW$, and the optimization (\ref{eq:def2}) reduces to optimizing over such state. This, in turn, is precisely the setup of Fig. \ref{fig:nnjn}(b), thus coinciding with our ansatz and proving its validity for pure $\rho_A$. As additional support for generic $\rho_A$, consider the relative entropy $S(\rho_{\fB}|\rho_{\fB0})$ as defined in eq. (\ref{eq:def3}). One can easily see that, for any ancilla-system coupling $V$, the corresponding value of $S(\rho_{\fB}|\rho_{\fB0})$ can be reproduced by a suitable ansatz with ancilla state $|\psi_{\fW}\rangle$. Of course, to prove that the ansatz leads to a global maximum one would need to verify this statement for the entire expression in (\ref{eq:def3}).

It is easy to see that (\ref{eq:ansatz}) does not recover the optimum in (\ref{eq:def2}) when $\fA$ and $\bar{\fA}$ are entangled. In Appendix \ref{app:entangled} we show an explicit counterexample.

We now show how the above ansatz leads to a self-consistent equation for the state on the ancilla $\fW_2$. We shall restrict to the single-replica case $N=1$ for simplicity.  Below, $\mathcal N:\fA\to \fB$ will denote the quantum channel obtained from the evolution of the total system after tracing out $\bar{\fB}$ and choosing $\rho_{\bar{\fA}}$ as the initial state for $\bar{\fA}$. Using the ansatz, the relative entropy in (\ref{eq:def2}) for $N=1$ specifies to:
\be\begin{split}\label{relent2} S(\rho_{\fontH {BW}}|\rho_{\fontH {BW},0})&=-S(\mathcal N(\rho_{\fontH {AW}}))+S(\rho_{\fontH W})-\tr(\mathcal N(\rho_{\fonth{AW}})(\log\mathcal N(\rho_{\text{in}})\otimes \text{Id}_{\fonth W})\\
&=-S(\mathcal N(|\psi\rangle\langle\psi|))+S(\rho_{\fonth W})-\tr\rho_{\fonth W}\mathcal N^\dag\log \mathcal N(\rho_{\text{in}})
\end{split}\ee
where $\rho_{\fA\fW}$ denotes the joint state of system-ancilla after the swap has been applied, $\rho_{\fonth W}= \tr_{\fonth W_2}|\psi_{\fW}\rangle\langle\psi_{\fW}|$. The reduced state of $\fW_2$ in $|\psi_{\fW}\rangle$ is equivalent to $\rho_{\fW}$ by unitary conjugation. Since all quantities we discuss are invariant under applying a unitary operator to $\fW_2$, we can assume the reduced density operator of $\fW_1$ and $\fW_2$ to be the same $\rho_{\fW}$ without loss of generality. 

Writing the quantum channel $\mathcal N(\rho)=\sum_I K_I\rho K_I^\dag$, where $K_I$ are the Kraus operators satisfying $\sum_I K^\dag_I K_I=\text{Id}$, it is useful to view the indices $I$ as labeling states $|I\rangle_{\fE}$ of an auxiliary system $\fE$, and to introduce the unitary evolution operator
\begin{align}
\sum_IK_I\otimes \ket{I}_{\fontH{E}}:\mathcal{H}_{\fontH{A}}\rightarrow \mathcal{H}_{\fontH{B}}\otimes \mathcal{H}_{\fontH{E}}
\end{align}
Tracing over $\fontH{E}$ leads to the channel $\mathcal{N}$, and tracing over $\fontH{B}$ leads to the complement channel $\tilde{\mathcal{N}}$, with
\be\label{Gamma} \tilde {\mathcal N}(\rho_{\fW})=\sum_{IJ}\Gamma_{IJ}\,|I\rangle_{\fE}\langle J|_{\fE},\qquad \Gamma_{IJ}=\tr(K_I\rho_{\fW} K_J^\dag)\,.\ee
In this notation, the first term in (\ref{relent2}) reduces to the entropy of the ancilla's output state, so that we can express eq. (\ref{relent2}) in terms of only the reduced state $\rho_{\fW}$:
\be\label{relent3}
S(\rho_{\fontH {BW}}|\rho_{\fontH {BW},0})=-S(\tilde{\mathcal N}(\rho_{\fW}))+S(\rho_{\fonth W})-\tr\rho_{\fonth W}\mathcal N^\dag\log \mathcal N(\rho_{\text{in}})\,.\ee
The optimization problem (\ref{eq:def2}) then reduces to optimizing over $\rho_{\fW}$. The variation of (\ref{relent2}) with respect to $\rho_{\fW}$ can be written as
\begin{align}
    \delta S={\rm Tr}\kc{\delta\rho_{\fontH{W}}\kc{\tilde{\mathcal{N}}^\dagger\log\mathcal{\tilde{N}}\kc{\rho_{\fontH{W}}}-\log\rho_{\fontH{W}}-\mathcal{N}^\dagger\log\mathcal{N}\rho_{\rm in}}}
\end{align}
thus leading to a self-consistent equation for $\rho_{\fontH{W}}$:
\begin{align}\label{ssol}
    \rho_{\fontH{W}}=C\exp\kd{\tilde{\mathcal{N}}^\dagger\log\mathcal{\tilde{N}}\kc{\rho_{\fontH{W}}}-\mathcal{N}^\dagger\log\mathcal{N}\kc{\rho_{\rm in}}}
\end{align}
with $C$ a normalization constant. One needs to remember to verify if the restricted (to $V=SWAP$) maximization is also a maximum (or at least a saddle point) for the unrestricted problem.

We now verify that the above ansatz is a stationary point of the optimization problem. Keeping into account factorization of the initial state (\ref{eq:fact}), the variation of the relative entropy $S(\rho_{\fB\fW}|\rho_{\fB\fW,0})$ that is inside the sup argument in eq. (\ref{eq:def2}) with respect to an infinitesimal change of the unitary $V\to (\text{Id}+iT)V$, with $T$ an infinitesimal Hermitian operator acting on $\fA\fW$, is
\be\begin{gathered} \label{genvar}\delta S=i\tr\, \left(T[\rho_{\fontH {AW}},-{\mathcal N}^\dag(\log{\mathcal N}( \rho_{\fontH {AW}}))+\text{Id}_{\fontH A}\otimes\log \bar\rho_{\fontH {W}}+\mathcal N^\dag(\log\mathcal N(\rho_{\text{in}}))\otimes \text{Id}_{\fontH W}]\right)\ ,\end{gathered}\ee
where $\bar \rho_{\fonth W}=\tr_B \rho_{\fonth{BW}}$. We now plug in the ansatz by taking $\fW=\fW_1\otimes\fW_2$, setting $V$ to be the swap operator between $\fA$ and $\fW_1$, and initializing the ancilla in a generic state $|\psi\rangle$. As a consequence, $\bar \rho_{\fonth W}=\rho_{\text{in}}\otimes\rho_{\fonth W}$, where $\rho_{\fW}$ was defined around (\ref{relent2}), $\rho_{\fonth {AW}}=(|\psi\rangle\langle\psi|)_{\fonth {AW}_2}\otimes(\rho_{\text{in}})_{\fonth W_1}$, and the above variation simplifies to
\be \delta S=i\tr\, \left(T\left[\text{Id}_{\fonth W_1}\otimes\big[|\psi\rangle\langle\psi|,-{\mathcal N}^\dag(\log{\mathcal N}( |\psi\rangle\langle\psi|))+\log \rho_{\fontH {W}}\otimes \text{Id}_{\fontH {W}_2}+\mathcal N^\dag(\log\mathcal N(\rho_{\text{in}}))\otimes \text{Id}_{\fontH W_2}\big]\right]\right)\ ,\ee
where we used that $[|\psi\rangle\langle\psi|,\text{Id}\otimes\rho_{\fonth W}]=[|\psi\rangle\langle\psi|,\rho_{\fonth W}\otimes \text{Id}]$. Assuming that $\mathcal N$ is unitary, the first term in the commutator vanishes. Plugging in (\ref{triv1}), we find that that latter is a stationary point of $S$ with respect to the general variation (\ref{genvar}). With generic $\mathcal N$ instead, we plug in (\ref{ssol}) and find
\be \label{deltas1}\delta S=i\tr\, T\big(\text{Id}_{\fonth W_1}\otimes\big[|\psi\rangle\langle\psi|,-{\mathcal N}^\dag(\log{\mathcal N}( |\psi\rangle\langle\psi|))+\tilde {\mathcal N}^\dag\log\tilde{\mathcal N}(\rho_{\fonth W})\otimes \text{Id}\big]\big)\,.\ee
Noting that
\be\begin{gathered} |\psi\rangle\langle\psi|\mathcal N^\dag(\mathcal N(|\psi\rangle\langle\psi|))^n=(\Gamma^n)_{IJ}|\psi\rangle\langle\psi|K_J^\dag K_I\\
\mathcal N^\dag(\mathcal N(|\psi\rangle\langle\psi|))^n|\psi\rangle\langle\psi|=(\Gamma^n)_{IJ}K_J^\dag K_I|\psi\rangle\langle\psi|\,,
\end{gathered}\ee
we conclude that the commutator in (\ref{deltas1}) vanishes. We thus showed that for a generic quantum channel $\mathcal N$, (\ref{ssol}) is a stationary point. It would be interesting to prove whether the ansatz is a local or global minimum in the general case.

Finally, we note that if the quantum channel $\mathcal N: A\to B$ is unitary, the ansatz allows us to analytically solve the optimization problem (\ref{eq:def2}). In this case the ancilla's entropy is zero, $S(\tilde{\mathcal N}(\rho_{\fW}))=0$, and eq. (\ref{ssol}) reduces to
\be\label{triv1} \rho_{\fonth W}=\frac{\rho_{\text{in}}^{-1}}{\tr \rho_{\text{in}}^{-1}}\ .\ee
Plugging the solution in the relative entropy (\ref{relent2}) leads to the space-time mutual information
\begin{align}\label{J1triv}
    J_1(\fA:\fB)=\log{\tr \rho_{\text{in}}^{-1}}\,.
\end{align}
This indicates that, when the initial state is pure, the state of the ancilla can be chosen so that $J_1(A:B)$ diverges, even when the Hilbert spaces of $\fA$ and $\fB$ are finite-dimensional. As we commented around eq. (\ref{eq:def3}), $J_1$ is unbounded, unlike the standard mutual information.


\section{Examples}\label{sec:examples}
In this section we shall study the behavior of the space-time mutual information in various quantum systems. We will first study its behavior in single-qubit systems subject to two different types of evolution and contrast the different behavior of the STMI in these two situations. We will then focus on many-body systems in two extreme cases: fully thermalizing and many-body localizing dynamics.

\subsection{Single-qubit system}

As a first example we consider the case where $\fA$ is the entire system and is given by a single qubit. The time evolution is defined by a quantum channel $\mathcal N$ which maps it to an output state in $\fB$, also a single qubit. We will consider two types of quantum channel: depolarizing channel $\mathcal N_{\text{dpl}}$ and dephasing channel $\mathcal N_{\text{dph}}$, defined by
\be \mathcal N_{\text{dpl}}(\rho)=(1-p)\rho+\frac p 2\text{Id},\qquad \mathcal N_{\text{dph}}(\rho)=\left(1-\frac p2\right)\rho+\frac p2\sigma_3\rho \sigma_3\,,\ee
with $0\leq p\leq 1$. Here $\sigma_i$ are the Pauli matrices. We will focus on the STMI (\ref{eq:def2}) with a single replica of the system $N=1$, i.e. $J_1(\fA:\fB)$. 

\begin{figure}
    \centering
    \includegraphics[width=5.8in]{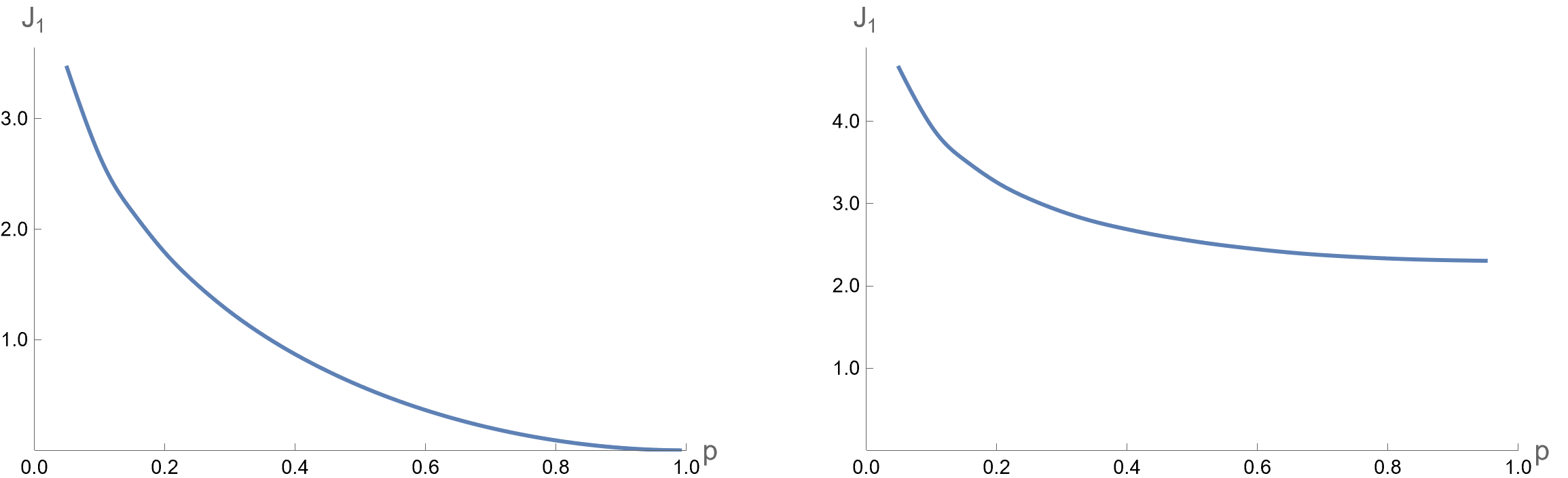}
    \caption{Plot of the STMI as a function of $p$ for the depolarizing channel $\mathcal N_{\text{dpl}}$ with initial state $\rho_{\text{in}}=|0\rangle\langle 0|$ (left) and for the dephasing channel $\mathcal N_{\text{dph}}$ with $\rho_{\text{in}}=\sqrt{1-\varepsilon}|0\rangle\langle0|+\sqrt\varepsilon|1\rangle\langle 1|$ and $\varepsilon\ll 1$ (right). For the dephasing channel, the STMI asymptotes to a finite value meaning that some information is preserved.}
    \label{fig:relent11}
\end{figure}

To numerically optimize over $V$, we initialize the state of $\fA\otimes \fW_1\otimes \fW_2$ to be $\rho_{\fA\fW}=\tilde V\rho_{\text{in}}\tilde V^\dag$, where $\tilde V: \fA\to \fA\otimes \fW_1\otimes \fW_2$ is a random isometry, and perform the following iterative updates on $\rho_{\fA\fW}$:
\be \rho_{\fA\fW}\to e^{-iM\eta}\rho_{\fA\fW}e^{iM\eta},\qquad M=iV\frac{\delta S}{\delta V}\,,\ee
where $\eta$ is a positive number that sets the increment. The explicit expression of $V\frac{\delta S}{\delta V}$ can be obtained from (\ref{genvar}).\footnote{The infinitesimal change $\delta V=V'-V$ is related to $T$ defined in (\ref{genvar}) through $V'=(\text{Id}+iT)V$, so that $V\frac{\delta S}{\delta V}=-i\frac{\delta S}{\delta T}$.} Fig. \ref{fig:relent11} shows the plots of $J_1(\fA:\fB)$ for the depolarizing and the dephasing channels as a function of $p$. Close to $p=0$ the evolution approaches the identity, and we recover (\ref{J1triv}). As $p\to 1$, $\mathcal N_{\text{dpl}}$ becomes a fully depolarizing channel; all the information from the past is lost and the space-time mutual information approaches zero. On the other hand, as $p\to 1$ the dephasing channel $\mathcal N_{\text{dph}}$ still preserves classical information of the initial state, and therefore the space-time mutual information approaches a nonzero constant.

We shall now analytically evaluate the STMI  $J_1(\fA:\fB)$ in a particular limit, using the ansatz of Sec. \ref{sec:ansatz}. We focus first on the dephasing channel $\mathcal N_{\text{dph}}$. We write the initial state as
\be\label{bloch} \rho_{\text{in}}=\frac 12 \left(\text{Id}+\vec a\cdot \vec \sigma\right)\ee
where $\vec a\in\mathbb R^3$ is a Bloch vector, with $|\vec a|= 1$, and we take $a_1=\vep$ and $a_3=\sqrt{1-\vep^2}$ with $\vep$ small, i.e., the state is almost an eigenstate of the evolution. In this case, the dephasing channel is close to the identity and we thus expect the space-time mutual information to diverge with $\vep$. Adopting the ansatz (\ref{eq:ansatz}), we take expression (\ref{relent3}) as our starting point and we optimize it over the reduced state of the ancilla $\rho_{\fW}$, which we parameterize using (\ref{bloch}) using the Bloch vector $\vec b$. We shall be interested in obtaining only the divergent part and it thus suffices to keep only the last term in (\ref{relent3}):
\be\label{relent4} S(\rho_{\fB\fW}|\rho_{\fB\fW,0})\approx -\tr\,\mathcal N_{\text{dph}}(\rho_{\fW})\log \mathcal N_{\text{dph}}(\rho_{\text{in}})\,.\ee
From $\mathcal N_{\text{dph}}(\rho_{\text{in}})=\frac 12(\text{Id}+(1-p)a_1\sigma_1+a_3\sigma_3)$, and from the identity $\frac 12(\text{Id}+\tanh \beta\vec n\cdot\vec\sigma)=\frac{e^{-\beta \vec n\cdot \vec\sigma}}{2\cosh\beta}$, with $\vec n^2=1$, we find
\be \log\mathcal N\rho_{\text{in}}=2\log\vep|1\rangle\langle 1|\,,\ee
which leads to
\be J_1(\fA:\fB)\approx -\sup_{\rho_{\fW}}\big(2\log\vep \,\langle 1|\mathcal N_{\text{dph}}(\rho_{\fW})|1\rangle\big)=
-2\log\vep\,.\ee
We then find that the space-time mutual information diverges as the initial state approaches an eigenstate of the evolution, and the divergent contribution is independent of $p$. We will see in the next subsection that a similar behavior happens for many-body localized systems.

\begin{figure}
    \centering
    \includegraphics[width=5.8in]{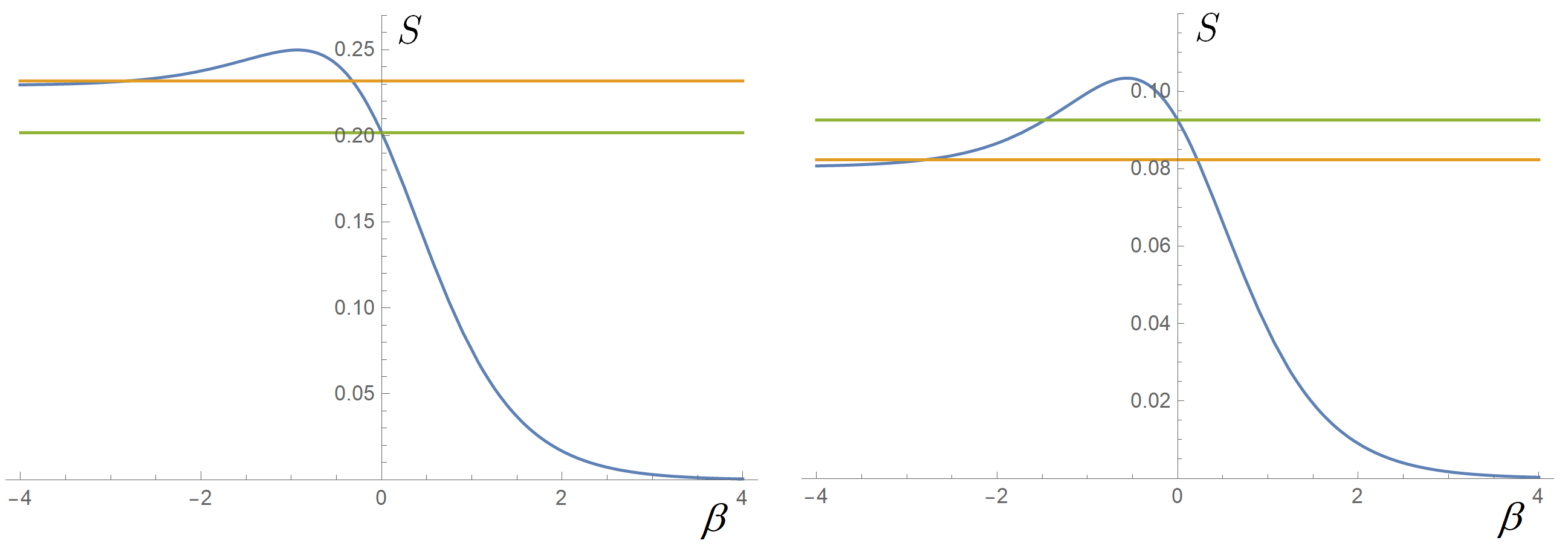}
    \caption{Plot of (\ref{relent2}) as a function of $\beta$ for a depolarizing channel with $p=0.5$ (left) and with $p=0.9$ (right). The STMI $J_1(\fA:\fB)$ is given by the maximum value of the blue curve. The green line represents the relative entropy using superdensity operator entanglement ($\beta=0$), while the orange line represents the entropy obtained by substituting (\ref{triv1}) in (\ref{relent2}). Depending on $p$, one quantity can be better than the other.}
    \label{fig:relent2}
\end{figure}

For the depolarizing channel $\mathcal N_{\text{dpl}}$ we consider a pure initial state, $\rho_{\text{in}}=|0\rangle\langle 0|$. Using again the ansatz (\ref{relent3}) we can exploit the symmetry of the evolution and choose the ancilla state
\be \rho_{\fW}=\frac{e^{\beta \sigma_3}}{2\cosh\beta}=\frac 12\text{Id}+\frac 12 \tanh\beta\sigma_3\,,\ee
and we expect eq. (\ref{ssol}) to be solved by $\rho_{\fW}\propto c_1\text{Id}+c_2\sigma_3$. The complementary channel $\tilde {\mathcal N}$ given in terms of matrix $\Gamma$ defined in (\ref{Gamma}) is\footnote{We are using the Kraus decomposition of the depolarizing channel: $\mathcal N_{\text{dpl}}(\rho)= \alpha_0^2 \rho + \sum_i \alpha_i^2\sigma_i\rho\sigma_i$.}
\be \label{gamma1}\Gamma=\begin{pmatrix}\alpha_0^2 & \alpha_0\alpha_jn_j\tanh\beta\\
\alpha_0 \alpha_i n_i\tanh \beta & \quad\alpha_i\alpha_j(\delta_{ij}-i\tanh \beta \varepsilon_{ijk}n_k)\end{pmatrix}\,,\ee
where $n_i=(0,0,1)$, $\vep_{ijk}$ is the totally antisymmetric tensor with $\vep_{123}=1$, and
\be\label{depo}\alpha_0=\sqrt{1-\frac 34p},\qquad \alpha_i=\sqrt{\frac p 4}\ .\ee
Using these expressions, we evaluate $J_1(\fA:\fB)$ by extremizing (\ref{relent3}) over $\beta$. A plot showing the relative entropy (\ref{relent2}) as a function of $\beta$ for various values of $p$ is given in Fig. \ref{fig:relent2}. The plot also illustrates a comparison between the STMI, the superdensity operator mutual information described at the end of Sec. \ref{sec:def} (corresponding to $\beta=0$), and the mutual information obtained by choosing the ancilla state (\ref{triv1}). 

We now find the exact value of the STMI for the depolarizing channel in the limit $p\to 1$. Using (\ref{relent3}),
\be\label{eq:dec1}
S(\rho_{\fontH {BW}}|\rho_{\fontH {BW},0})=-S(\tilde{\mathcal N}(\rho_{\fW}))+S(\rho_{\fonth W})-\tr\mathcal N(\rho_{\fonth W})\log \mathcal N(\rho_{\text{in}})\,.\ee
It is straightforward to evaluate each term in (\ref{eq:dec1}). As $p\to 1$, we obtain
\be\label{eq:dec2}
S(\rho_{\fontH {BW}}|\rho_{\fontH {BW},0})=
A(1-p)^2 + B(1-p)^3+\cdots\ee
where
\be\begin{split}
A&=e^{-2\beta}(1+\text{coth}\beta)(1+\beta+\beta\,\text{coth}\beta)\tanh\beta\\
B&=-\beta\,\text{csch}\beta\,\text{sech}\beta\,.
\end{split}\ee
Numerically solving the optimization (\ref{eq:def2}) for this relative entropy, we find that $\beta\approx -0.72-(1-p)0.68$, and we thus see that $J_1(\fA:\fB)\to 0$ linearly as $1-p\to 0$. In the fully depolarizing case $p=1$ the optimization problem is trivial as $\fA$ and $\fB$ are disconnected, thus any ancilla state $\rho_{\fW}$ solves the optimization problem. However, what we find shows that the limit $p\to 1$ selects a unique ancilla state.

\subsection{MBL and Thermalization}

We will now explore two extreme cases of many-body dynamics. The first example, a many-body localized (MBL) system, preserves an extensive amount of local operators, while the second example concerns a thermalizing system, and thus all local information is efficiently scrambled across the system. These two examples thus constitute contrasting cases where the STMI should display very different phenomenology.

\begin{figure}
    \centering
    \includegraphics[width=5in]{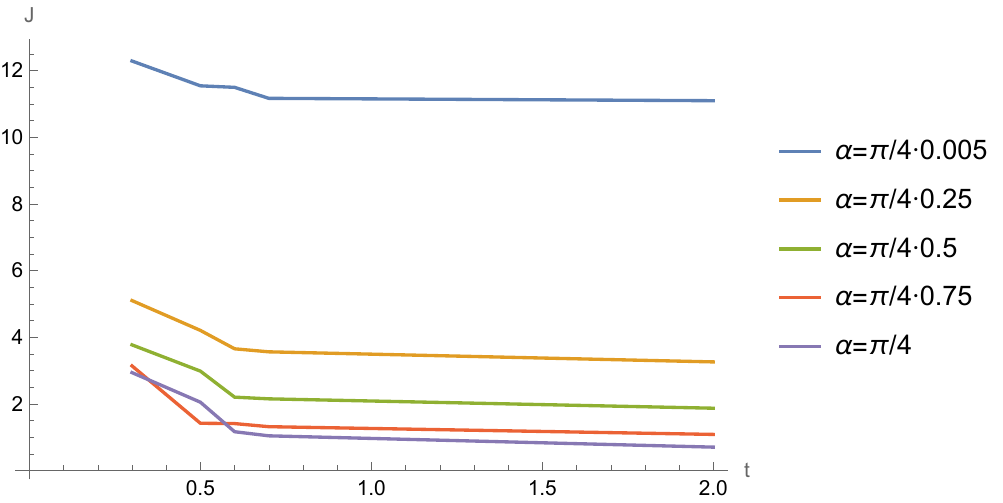}
    \caption{Plot of $J_1(\fA:\fB)$ for various values of $\alpha$ with evolution given by the MBL fixed point Hamiltonian (\ref{fpham}). 
    Each of these plots is obtained from a single disorder realization, with $w=10$ and $\xi=2$.}
    \label{fig:alpha}
\end{figure}

We shall start with MBL and study the single-replica STMI $J_1(\fA:\fB)$, with $\fA$ a single qubit at the initial time $t=0$ and $\fB$ the same qubit at time $t$. As a model we consider a truncation of the MBL fixed-point Hamiltonian \cite{serbyn2013local,huse2014phenomenology,chandran2015constructing}:
\be\label{fpham} H=\sum_i h_i \sigma^3_i+\sum_{i<j}J_{ij}\sigma_i^3\sigma_j^3+\sum_{i<j<k}J_{ijk}\sigma_i^3\sigma_j^3\sigma^3_k\,,\ee
where the Pauli matrix on each site $\sigma_i^3$ is a local conserved operator. Here, $J_{ij}=e^{-|i-j|/\xi}\tilde J_{ij}$ and $J_{ijk}=e^{-|i-k|/\xi}\tilde J_{ijk}$, and $h_i,\tilde J_{ij},\tilde J_{ijk}\in[-w,w]$ are drawn from a uniform distribution. Assuming the initial state is factorized between $\fA$ and $\bar {\fA}$, tracing over $\bar {\fA}$ gives a channel $\mathcal N:\fA\to \fB$. We consider initial pure states on $\fA$ of the form $\rho_{\text{in}}=|\chi\rangle\langle\chi|$, where
\be \label{chiin}|\chi\rangle=\cos\alpha|0\rangle+\sin\alpha|1\rangle\,.\ee
It is easy to see that states $|0\rangle,|1\rangle$ are preserved by the evolution. This means that, writing in matrix form using the $|0\rangle,|1\rangle$ basis, $\mathcal N$ must act as
\be \mathcal N\rho_{\text{in}}=\begin{pmatrix}
\cos^2\alpha&f(t)\cos\alpha\sin\alpha \\ f^*(t)\cos\alpha\sin\alpha & \sin^2\alpha\end{pmatrix}\ee
where $f(0)=1$ and, as $t$ grows, we expect generically that $f(t)$ vanishes, similarly to the fully dephasing channel $\mathcal N_{\text{dph}}$ discussed in the previous subsection. If $\alpha=0$, the state does not evolve and remains pure, so that $J_1(\fA:\fB)=\infty$ for all times, consistently with the discussion around (\ref{J1triv}). For general $\alpha$, and because $\rho_{\text{in}}$ is pure, we can decompose the relative entropy $S(\rho_{\fB\fW}|\rho_{\fB\fW,0})$ as in (\ref{relent3}):
\be\label{Srho1}
S(\rho_{\fontH {BW}}|\rho_{\fontH {BW},0})=-S(\tilde{\mathcal N}(\rho_{\fW}))+S(\rho_{\fonth W})-\tr\left(\mathcal N\rho_{\fonth W}\log \mathcal N(\rho_{\text{in}})\right)\,.\ee
For small $\alpha$, the first two terms in (\ref{Srho1}) are finite, while the last one diverges. Applying the same manipulations as those around eq. (\ref{relent4}), we find that
\be\label{J1mbl} J_1(\fA:\fB)\approx -2\log\alpha\,,\ee
showing that the closer the initial state (\ref{chiin}) is to an eigenstate of the conserved operator $\sigma^3$, the larger the STMI will be at late time, consistently with the intuition that the STMI quantifies the information preserved by the system. Fig. \ref{fig:alpha} shows the time-dependence of $J_1(\fA:\fB)$ and confirms this behavior.

\begin{figure}
    \centering
    \includegraphics[width=5in]{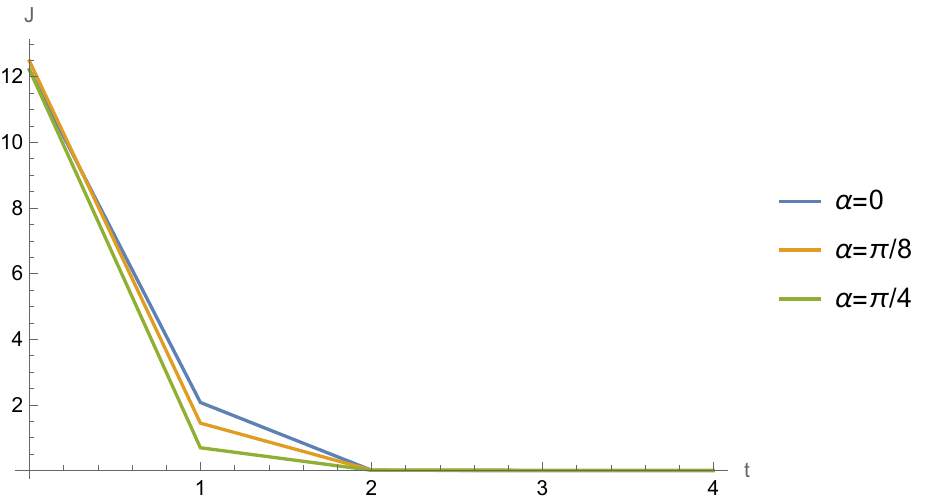}
    \caption{Plot of $J_1(\fA:\fB)$ for the Floquet evolution generated by (\ref{floq}). The initial state is $\rho_{\text{in}}=(1-\vep)|\chi\rangle\langle\chi|+\frac\vep 2\,\text{Id}$, with $|\chi\rangle$ given in (\ref{chiin}), and where $\vep=10^{-5}$ is used to regularize the initial value of the STMI. }
    \label{fig:thermal}
\end{figure}

For the thermalizing case, we consider a Floquet system whose evolution is generated by the unitary
\be\label{floq}\begin{gathered} U=e^{-i\frac \tau 2 H_x}e^{-i\tau H_z}e^{-i\frac \tau 2 H_x}\\
H_x=g\sum_{j=1}^L \sigma^x_j,\qquad H_z=\sum_{j=1}^{L-1}\sigma_j^z\sigma_{j+1}^z+h\sum_{j=1}^L
\sigma_j^z\,.\end{gathered}\ee
For $(g,h,\tau)=(0.9045,0.8090,0.8)$, this system is known to thermalize efficiently and have weak finite-size effects \cite{zhang2016floquet}. We consider the same STMI $J_1(\fA:\fB)$ as in the MBL case with coincident location of input and output qubits. Since there are no conserved operators in this case (including energy), we expect that $J_1(\fA:\fB)$ will quickly drop to zero for any initial state. Fig. \ref{fig:thermal} shows $J_1(\fA:\fB)$ for various initial states, confirming our expectation. The timescale characteristic of this drop is given by the rate of decoherence caused by the effective channel describing the evolution of a single qubit. 

\section{Classical space-time mutual information}\label{sec:class}
\begin{figure}
    \centering
    \includegraphics[width=6in]{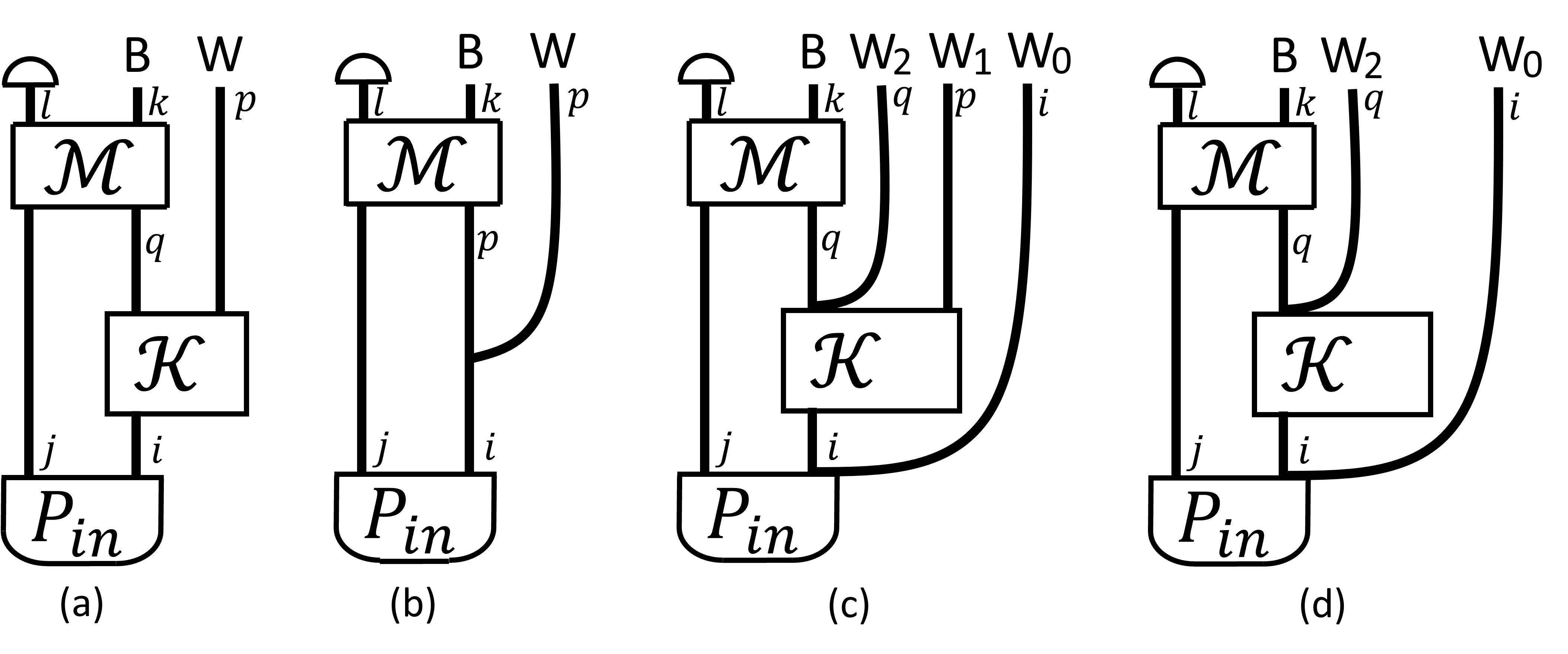}
    \caption{Representations of the probability $P_{\fB\fW}$ for various ancilla-system couplings.}
    \label{fig:classic}
\end{figure}

We now discuss how the STMI can be defined for classical systems, and make connection to information-theoretic quantities discussed in earlier literature. Consider a classical system whose initial state is characterized by a probability over a state space $\fontH S$, which we denote by $P_{\text{in}}(i)$, where $i$ labels a state in $\fontH S$. This probability is then mapped to an output probability through a stochastic map $\mathcal M:\ \fontH S\to \fontH S'$ acting as $P_{\text{in}}(i)\to \sum_i \mathcal M(j|i)P_{\text{in}}(i)$, where $j$ labels states of $\fontH S'$, and $\mathcal M(j|i)$ denotes the transfer matrix associated to the map $\mathcal M$ that satisfies normalization and positivity: $\sum_j \mathcal M(j|i)=1$ and $\mathcal M(j|i)\geq 0$. The notion of STMI also applies to this classical setting with the important difference that the ancilla, as well as its coupling to the system, are themselves restricted to be classical. With $\fA\subseteq \fontH S$ and $\fB \subseteq \fontH{S}'$, we introduce a classical ancilla $\fontH{W}$ and a stochastic map $\mathcal K$ that acts jointly on $\fA\fontH{W}$. From this we can define connected and disconnected ancilla-system probabilities, similarly to Sec. \ref{sec:defi}. Taking a single copy of the system $N=1$, these probabilities are:
\be \begin{split} \label{PWclass} P_{\fB\fW}(kp)&=\sum_{ljqri}\mathcal M(kl|qj)\mathcal K(qp|i)P_{\text{in}}(ij)\\
P_{\fB\fW,0}(kp)&=P_{\fB,0}(k)P_{\fW}(p)
\,,\end{split}\ee
where $P_{\fB,0}(k)=\sum_{ijl}\mathcal M(kl|ij)P_{\text{in}}(ij)$ is the unperturbed output state and $P_{\fW}(p)=\sum_k P_{\fB\fW}(kp)$. Here, $k,l,p,i$ and $j$ label states in $\fB,\bar{\fB},\fW,\fA$ and $\bar{\fA}$ respectively. We represented the state for generic $N\geq 1$ in Fig. \ref{fig:classic}(a). The classical counterpart of the STMI is then
\be\label{def:class} J_N(\fA:\fB)=\frac 1N\sup_{\mathcal K} D(P_{\fB^N\fW,1}|P_{\fB^N\fW,0})\,,\ee
where $D(\cdot|\cdot)$ denotes the Kullbeck-Leibler (KL) divergence, and we assume $\fW$ to be sufficiently large.

One can show that, similarly to the quantum STMI, (\ref{def:class}) bounds correlation functions as well as response functions. In fact, correlation functions can be already bounded by the input-output mutual information,\footnote{By intput-output mutual information, we simply mean the mutual information between $A$ and $B$, where these have joint probability distribution $p(ki)=\sum_{lj}\mathcal M(kl|ij)P_{\text{in}}(ij)$ \cite{cover2012elements}.} which can be obtained from the KL divergence in (\ref{def:class}) by choosing $\mathcal K$ to be the copy channel: $\mathcal K(qp|i)=1$ if $q=p=i$ and 0 otherwise, see Fig. \ref{fig:classic}(b). To see this, we write the correlation function of two
observables $\mathcal O_{\fA}$ and $\mathcal O_{\fB}$ as
\be\label{2ptcl} \langle \mathcal O_{\fB}(t)\mathcal O_{\fA}\rangle=\sum_{ijkl}{\mathcal M}(kl|ij)P_{\text{in}}(ij)
\mathcal O_{\fB}(k)\mathcal O_{\fA}(i)\,.\ee
Using similar steps as in Sec. \ref{sec:bound}, one easily obtains the bound 
$I(\fB:\fA)\geq \frac 12
\left|\frac{\langle \mathcal O_{\fB}(t)\mathcal O_{\fA}\rangle_c}{||\mathcal O_{\fB}||_{\infty}||\mathcal O_{\fA}||_{\infty}}\right|^2$, where $||\mathcal O||_{\infty}=\sup_i|\mathcal O(i)|$.
We then see that, in the classical case, correlations can be bound without adaptively optimizing over the system-ancilla coupling.

To bound response functions, on the other hand, the input-output mutual information is not enough: we need an adaptive ancilla-system coupling, like in the quantum case. Consider perturbing the evolution before applying channel $\mathcal M$ through a small perturbation from the identity, which itself can be viewed as a channel. Its transfer matrix can be written as $\mathcal N_{\fA}(k|i)=\delta_{ki}+\vep N_{\fA}(k|i)$, with $\sum_k N_{\fA}(k|i)=0$, $N_{\fA}(k|i)\geq 0 $ for $k\neq i$ and $\vep\geq 0$ small. The response function is then the leading order contribution in $\vep$ to the one-point function of $\mathcal O_{\fB}$:
\be\begin{split} G_{R}(\mathcal O_{\fB},N_{\fA})&=\lim_{\vep\to 0}\vep^{-1}\left.\left(\langle \mathcal O_{\fB}(t)\rangle_{\mathcal N_{\fA}}-\langle \mathcal O_{\fB}(t)\rangle_{\mathcal N_{\fA}=\text{Id}}\right)\right.\\
&=
\sum_{jiklp}\mathcal O_{\fB}(j)\mathcal M(jl|kp)N_{\fA}(k|i) P_{\text{in}}(ip)\,.\end{split}\ee
The input-output mutual information will not in general bound such two-point functions. For example, take $P_{\text{in}}(i)=\delta_{i0}$ to be a pure state, then $I(\fA:\fB)=0$. One can easily see that there are choices of $\mathcal O_{\fB}$ and $N_{\fA}$ with a nonzero response function so that the bound is violated.

We shall now prove the bound using $J_1(\fA:\fB)$, adopting a similar approach as in Sec. \ref{sec:bound}. Introduce a 2-bit ancilla with states $p=0,1$, and the ancilla-system coupling
\be \mathcal K(j0|i)=\frac 12\left(\delta_{ij}+\frac{N_A(j|i)}{||N_A||_{\infty}}\right),\qquad \mathcal K(j1|i)=\frac 12 \delta_{ij}\,,\ee
where $||N_A||_\infty=\sup_j\sum_i|N_A(i|j)|$. Note that $\mathcal K$ satisfies positivity and normalization. Further introducing the observable $\mathcal O_{\fW}$ acting on the ancilla $\fW$, with $\mathcal O_{\fW}(p=0)=1=-\mathcal O_{\fW}(p=1)$, we find
\be \sum_{ipkj} \mathcal O_{\fB}(j)\mathcal O_{\fW}(p)\mathcal M(jl|k)\mathcal K(kp|i)P_{\text{in}}(i)=\frac 12\frac{G_R(\mathcal O_{\fB},N_{\fA})}{||N_{\fA}||_{\infty}}\,,\ee
where for simplicity we left implicit the dependence on the indices in $\bar{\fB}$ and $\bar {\fA}$. Applying similar steps as in Sec. \ref{sec:bound} one then shows that the desired bound holds $J_1(\fA:\fB)\geq \frac 18 \left(\frac{G_R(\mathcal O_{\fB},N_{\fA})}{||\mathcal O_{\fB}||_\infty||N_{\fA}||_{\infty}}\right)^2$, where the factor of $\frac 18$ also appeared in Theorem \ref{thm} for the quantum case.

Finally, we note that the expression of the classical STMI can be slightly simplified. Indeed, still taking to $N=1$ for simplicity, we now show that to find the optimal $J_1$ one can restrict the state $P_{\fB\fW}$ in (\ref{PWclass}) to the form 
\be \label{PWred} P_{\fB\fW}(kqi)=\sum_{lj}\mathcal M(kl|qj)\mathcal K(q|i)P_{\text{in}}(ij)\ee
as illustrated in Fig. \ref{fig:classic}(d). The key is that copying to the ancillas $\fW_0$ and $\fW_2$ the state before and after applying $\mathcal K$, as in Fig. \ref{fig:classic}(c), does not require introducing any additional probing of the system, i.e. $\fW_1$ is not necessary. More precisely, the conditional mutual information $I(\fB:\fW_1|\fW_0\fW_2)$ can be readily shown to vanish. Due to this fact, and using the classical counterpart of (\ref{eq:def3}), we have
\be D(P_{\fB\fW}|P_{\fB\fW,0}) = I(\fB:\fW_0\fW_2)+D(P_{\fB}|P_{\fB,0})\,,\ee
thus showing that marginalizing over $\fW_1$ does not affect $D(P_{\fB\fW}|P_{\fB\fW,0})$, and the optimal $\mathcal K$ in (\ref{def:class}) can be achieved using (\ref{PWred}). A consequence of this fact is that, if the initial state is factorized, i.e. $I(\fA:\bar{\fA})=0$, from Fig. \ref{fig:classic}(d) it is clear that maximization over $\mathcal K$ can be replaced by a maximization over the state of $\fA$ with $\mathcal K$ fixed to the identity.

While we are not aware of discussions of the STMI in the literature, a restricted version of our implementation has been considered in the context of classical channel discrimination, where one optimizes over the input state \cite{hayashi2009discrimination,harrow2010adaptive}.

\section{Conclusions}\label{sec:concl}
In this paper, we introduced the space-time mutual information (STMI), a quantity that generalizes mutual information to spatial subregions that can be separated in time. This was achieved by demanding that the STMI satisfies some of the natural properties possessed by the standard mutual information. The most stringent property leading to our definition (\ref{eq:def2}) is that the STMI should bound space-time correlation functions between the two subregions. 

We then investigated several properties that descend from our proposal, such as the Markov property and the relationship to quantum channel discrimination. We studied the behavior of the STMI in MBL and thermalizing many-body systems and found very distinct behaviors, thus, in a sense, providing a characterization of these two types of dynamics. Finally, we discussed a classical counterpart of the STMI. 

Our framework can be extended in several directions. First, in this work we studied the time dependence of the STMI for two extreme cases in the context of many-body dynamics (MBL and Floquet thermalization). A natural next step is to look at more intermediate situations, e.g. thermalizing systems conserving a finite number of quantities such as energy or charge, or kinematically constrained models \cite{glorioso2022breakdown,richter2022anomalous,singh2021subdiffusion}. For subregions small enough compared to system size we expect the STMI to decay polynomially in time for these systems. When the subregions considered become large, we saw around eq. (\ref{asymptfin}) that the STMI asymptotes to a finite value; it would be interesting to find how this asymptotic value is approached at late times. Another intriguing avenue for investigation involves examining the time dependence of the STMI as a diagnostic tool to differentiate between integrable and non-integrable systems, as explored in recent studies such as those highlighted in \cite{lerose2021scaling,giudice2022temporal}.
Additionally, it will also be interesting to consider restrictions of the optimization over $V$ which may characterize the type of information that the ancilla is able to extract from the system. For example, one can restrict $V$ to be a one-way LOCC (local operations and classical communication) from $A$ to $W$, which corresponds to an experimentalist who can only carry out classical measurements.

On an information-theoretic level, it is still an open question whether the STMI satisfies additivity. A positive answer to this would imply that it is sufficient to restrict to a single replica $N=1$ in the definition of the STMI (\ref{eq:def2}). We proved additivity in a restricted case where we could map our quantity to the channel relative entropy. While we could not find counter-examples to additivity in more general settings, it is still possible that additivity might not hold in full generality; we leave this question to future work.

We thank Aditya Cowsik, Patrick Hayden, Anton Kapustin, Vedika Khemani and Jinzhao Wang for illuminating discussions. This work is supported
by the Department of Energy through Award
DE-SC0019380 (PG and XLQ), the National
Science Foundation under grant No.2111998 (XLQ), the It from Qubit Simons Collaboration (ZY), the Simons Fundation through Award No. 560571 (XLQ), the CAREER Award DMR-2045181 (PG), the Laboratory for Physical Sciences through
the Condensed Matter Theory Center of the University of Maryland, College Park (PG), and the Simons Foundation through Award No. 620869 (PG). We are grateful to the long term workshop YITP-T-23-01 held at YITP, Kyoto University, where a part of this work was done. XLQ is grateful for the hospitality of Institute for Advanced Study, Tsinghua University (IASTU), where he visited while carrying out part of this work.

\appendix

\section{Bound on correlations with superdensity operator}\label{app:bound}
Here we obtain a bound on correlations similar to (\ref{eq:ret}),(\ref{eq:sym}) using the relative entropy
$S\left(\rho_{\fontH{B}\fontH{W}}\middle|\rho_{\fontH{B},0}\otimes\rho_{\fontH{W}}\right)$, where $V=S_{\fA\fW_1}\otimes\mathbb 1$ is the swap between $\fA$ and $\fW_1$, and the ancilla $\fW_1\fW_2$ is prepared in an EPR state (see also Fig. \ref{fig:def1}). Using the norm bound,
\begin{align}
    S\left(\rho_{\fontH{B}\fontH{W}}\middle|\rho_{\fontH{B},0}\otimes\rho_{\fontH{W}}\right)\geq \frac12\norm{ \rho_{\fontH{B}\fontH{W}_1\fontH{W}_21}-\rho_{\fontH{B}\fontH{W}_1\fontH{W}_20}}_1^2\geq \frac{\kc{\avg{{\mathcal O}_{\fontH{B}}{\mathcal O}_{\fontH{W}}}_{\text c}}^2}{2\norm{\mathcal O_{\fontH{B}}}_\infty^2\norm{\mathcal O_{\fontH{W}}}_\infty^2}\label{eq:corr bound}
\end{align}
for any Hermitian operators $\mathcal O_{\fontH{B}},\mathcal O_{\fontH{W}}$. The norm in the denominator is the operator norm, {\it i.e.} absolute value of the maximal eigenvalue, and where $\langle \mathcal O_{\fB}(t)\mathcal O_{\fW}\rangle_{\text c}$ denotes any connected Schwinger-Keldysh time ordered two-point function. For example, consider a traceless Hermitian operator $O_{\fontH{A}}$, and define an operator on $\fontH{W}$ by
\begin{align}
    \mathcal O_{\fontH{W}}=i\kc{\ket{I}\bra{I}O_{\fontH{A}}-O_{\fontH{A}}\ket{I}\bra{I}}
\end{align}
with $\ket{I}$ the initial state of $\fontH{W}$, and $O_{\fontH{A}}$ acts on $\fontH{W}_1$. When $O_{\fontH{A}}$ is traceless, $\bra{I}O_{\fontH{A}}\ket{I}=0$, and the norm of $\mathcal O_{\fontH{W}}$ satisfies
\begin{align}
    \norm{\mathcal O_{\fontH{W}}}_\infty\leq 2\norm {\mathcal O_{\fA}}_{\infty}
\end{align}
On the other hand, inserting $\mathcal O_{\fontH{W}}$ into Eq. (\ref{eq:corr bound}) leads to
\begin{align}\label{2ptqt}
    \avg{\mathcal O_{\fontH{B}}\mathcal O_{\fontH{W}}}_{\text c}=d_{\fontH{A}}^{-2}i\,{\rm tr}\kc{\kd{\mathcal O_{\fontH{B}}(t),O_{\fontH{A}}}\rho_{\rm in}}\,.
\end{align}
Therefore we obtain an upper bound of response functions:
\begin{align}\label{Sboun}
    S\left(\rho_{\fontH{B}\fontH{W}}\middle|\rho_{\fontH{B},0}\otimes\rho_{\fontH{W}}\right)\geq\frac{\kc{i\,{\rm tr}\kc{\kd{\mathcal O_{\fontH{B}}(t),O_{\fontH{A}}}\rho_{\rm in}}}^2}{8d_{\fontH{A}}^4\norm{\mathcal O_{\fontH{B}}}_\infty^2\norm{O_{\fontH{A}}}_\infty^2}
\end{align}
Note that the left-hand side can be replaced by the mutual information term $I(\fontH{B}:\fontH{W})$ (the super-density operator mutual information \cite{cotler2018superdensity}) because the disconnected part of the correlator vanishes, $\avg{\mathcal O_{\fontH{W}}}=0$. The tighter bound is then
\begin{align}\label{miboun}
    I(\fontH{B}:\fontH{W})\geq\frac{\kc{i\,{\rm tr}\kc{\kd{\mathcal O_{\fontH{B}}(t),O_{\fontH{A}}}\rho_{\rm in}}}^2}{8d_{\fontH{A}}^4\norm{\mathcal O_{\fontH{B}}}_\infty^2\norm{O_{\fontH{A}}}_\infty^2}\,.
\end{align}
Note that (\ref{Sboun}) and (\ref{miboun}) come with an additional factor of $d_{\fA}^{-4}$ compared to (\ref{eq:ret}) and (\ref{miboun1}), as stated in the main text, thus the bound we just proved is weaker. Similarly, one can obtain a bound for the symmetric two-point function (\ref{eq:sym}) suppressed by the same factor.

In concluding this Appendix, we note that we can bound the causal influence $CI(A:B)$ \cite{cotler2019quantum} by inserting a unitary $U_{\fA}$ in $\fA$. This corresponds to
\begin{align}
    \mathcal O_{\fontH{W}}=U_{\fontH{A}}\ket{I}\bra{I}U_{\fontH{A}}^\dagger
\end{align}
where in the above, $U_{\fontH{A}}$ acts on $\fontH{W}_1$. Note that $\norm{\mathcal O_{\fontH{W}}}_{\infty}=1$. 
The inequality (\ref{eq:corr bound}) becomes
\begin{align}
    S\left(\rho_{\fontH{B}\fontH{W}}\middle|\rho_{\fontH{B},0}\otimes\rho_{\fontH{W}}\right)\geq\frac{\kc{\avg{\mathcal O_{\fontH{B}}(t)}(U_{\fontH{A}})-\avg{\mathcal O_{\fontH{B}}(t)}(\mathbb{1}_{\fontH{A}})}^2}{2d_{\fontH{A}}^2\norm{\mathcal O_{\fontH{B}}}^2_{\infty}}\,.
\end{align}

\section{Additivity for general initial states}\label{app:addgeneral}


For generic $N$ the initial state is $\rho_{\text{in}}=\rho_{\text{in},\text{single\ copy}}^{\otimes N}$, and the state after coupling to the ancilla is $\rho_{(\bar{\fA}\fA)^N\fonth{W}}=V(\rho_{\text{in}}\otimes|\psi\rangle\langle\psi|)V^\dag$. Restricting to unitary evolution for simplicity, and denoting the evolution of the replicated system by $U_N=U^{\otimes N}$, the connected and disconnected final states are \be\label{rho01}\begin{split}\rho_{\fB^N\fW}=\tr_{\bar{\fB}^N}\,U_N
\rho_{(\bar{\fA}\fA)^N\fonth{W}}U_N^\dag\qquad
\rho_{\fB^N\fW0}=\left(\tr_{\bar{\fB}^N}\,U_N\rho_{\text{in}}U_N^\dag\right)
\otimes\rho_{\fW}\,,\end{split}\ee
where $\rho_{\fW}=\tr_{(\bar{\fA}\fA)^N}\,\rho_{(\bar{\fA}\fA)^N\fonth{
W}}$. Consider an infinitesimal variation of $V$, $V\to (\text{Id}+iT)V$, with $T$ a Hermitian matrix acting on $\fA^N\fW$. This gives $\delta\rho_{(\bar{\fA}\fA)^N\fonth{W}}=i[T,\rho_{(\bar{\fA}\fA)^N\fonth{W}}]$, so that
\be\begin{split}\delta S(\rho_{\fB^N\fonth{W}}|\rho_{\fB^N\fonth{W},0})=&
i\,\tr_{(\bar {\fA}\fA)^N\fonth{W}}T\,\left[\rho_{(\bar{\fA}\fA)^N\fonth{W}},M_N\right]\,,\end{split}\ee
where
\be\begin{split} \label{eqMN} M_N=&U_N^\dag\left(\text{Id}_{\bar{\fB}^N}\otimes\log\rho_{\fonth{B}^N\fW}\right)U_N-U_N^\dag\left(\text{Id}_{\bar{\fB}^N}\otimes\log\left(\tr_{\bar{\fB}^N}U_N\rho_{\text{in}}U_N^\dag\right)\right)U_N\otimes\text{Id}_{\fonth W}\\
&-\text{Id}_{(\bar{\fA}\fonth{A})^N}\otimes\log\rho_{\fW}\,.\end{split}\ee
Restricting to factorized ancilla-system coupling $V=V_1^{\otimes N}$, we have $\rho_{(\bar {\fA}\fA)^N\fonth{W}}=(\rho_{\bar{\fA}\fA\fonth{W}})^{\otimes N}$, and thus $M_N=M_1\otimes\text{Id}^{\otimes(N-1)}+\text{Id}\otimes M_1\otimes\text{Id}^{\otimes(N-2)}+\cdots$, where $M_1$ is the single-copy version of (\ref{eqMN}). We then find
\be\label{var3}\begin{split}\delta S(\rho_{\fB^N\fonth{W}}|\rho_{\fB^N\fonth{W},0})
=&i\,\tr_{(\bar{\fA}\fA)^N\fonth{W}}\,T\,\bigg([\rho_{\bar{\fA}\fonth{AW}},M_1]\otimes(\rho_{\bar{\fA}\fA\fonth{W}})^{\otimes(N-1)}\\
&+\rho_{\bar{\fA}\fA\fonth{W}}\otimes[\rho_{\bar{\fA}\fonth{AW}},M_1]\otimes(\rho_{\bar{\fA}\fA\fonth{W}})^{\otimes(N-2)}+\cdots\bigg)=0\ ,\end{split}\ee
where in the last step we used that $V_1$ is a stationary point for $S(\rho_{\fB\fW}|\rho_{\fB\fW,0})$, as this is equivalent to $[\rho_{\bar{\fA}\fonth{AW}},M_1]$. We thus showed that $V_N=V_1^{\otimes N}$ is a stationary point for the $N$-replica optimization (\ref{eq:def2}). This however does not imply that such $V_N$ is a global minimum, and thus we cannot conclude the additivity of $J_N$ for general initial states. We leave this as an open question for future work.

\section{Entangled initial state}\label{app:entangled}
\begin{figure}
    \centering
    \includegraphics[width=6in]{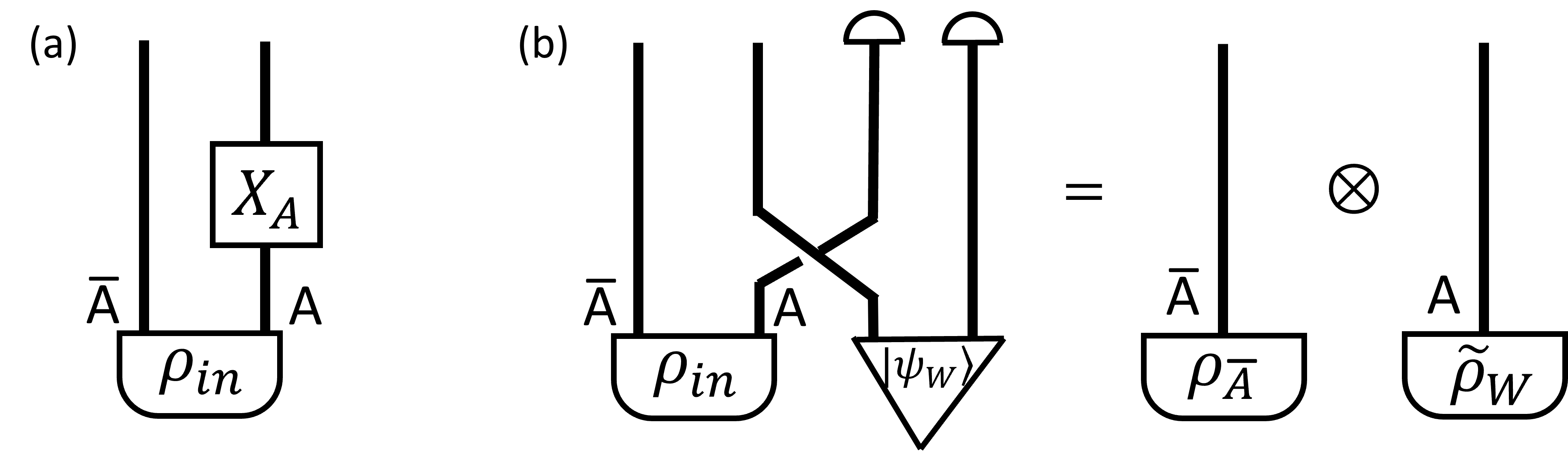}
    \caption{The setup in (a) yields a larger relative entropy compared to the one in (b) when $\rho_{\text{in}}$ is given by (\ref{rhoinapp}), thus providing a counterexample to the ansatz of Sec.\ref{sec:ansatz} when $\fA$ and $\bar{\fA}$ are entangled.}
    \label{fig:swapX}
\end{figure}

We here discuss a counterexample to the ansatz proposed in Sec. \ref{sec:ansatz} when $\fA$ and $\bar{\fA}$ are entangled. Suppose $\fA$ and $\bar{\fA}$ are two qubits, with an entangled initial state
\be \label{rhoinapp}\rho_{\text{in}}=(1-\vep)|\Gamma_{\bar {\fA}\fA}\rangle\langle \Gamma_{\bar {\fA}\fA}|+\frac \vep 4\text{Id}\,,\ee
where $|\Gamma\rangle$ denotes an EPR state. Let us take the evolution of the system to be trivial for simplicity: $\rho_{\fB0}=\rho_{\text{in}}$, with $\fB$ isomorphic to $\bar \fA\fA$. We first evaluate $J(\fA:\fB)$ by considering $V=X_A$ (i.e. the coupling to ancilla is the $X$ Pauli matrix acting on $\fA$: coupling to the ancilla $\fW$ is trivial and we take $\fW=\emptyset$), i.e.
\be \rho_B=\tr_W\rho_{B W}=X_A\rho_{\text{in}}X_A=(1-\vep)X_A|\Gamma_{\bar {\fA}\fA}\rangle\langle \Gamma_{\bar {\fA}\fA}|X_A+\frac \vep 4\text{Id}\,,\ee
see Fig. \ref{fig:swapX}(a). We have
\be\label{XJ} J(\fA:\fB)\geq S(\rho_{\fB}\rho_{\fB0})=-\log\vep+O(\vep\log\vep)
\ee
where in the last step we only kept the divergent part in $\vep$, and where we used
\begin{align} \tr \rho_{\fB}\log\rho_{\fB}&=
\left(1-\frac 34 \vep\right)\log\left(1-\frac 34\vep\right)+\frac 34 \vep\log\frac\vep4\\
-\tr \rho_{\fB}\log\rho_{\fB0}&=-\left(1-\frac 34\vep\right)\log\frac \vep 4-\frac \vep 4\log\left(1-\frac 34 \vep\right)-2\frac \vep 4\log\frac \vep 4\,.
\end{align}

Let us now estimate $J(\fA:\fB)$ using $V=$ swap, and placing $\fW$ in an arbitrary initial state. Then (see Fig. \ref{fig:swapX}(b))
\be \rho_{\fB}=\rho_{\bar \fA}\otimes \tilde \rho_{\fW},\quad \rho_{\bar \fA}=\frac 12 \text{Id},\quad \tilde \rho_{\fW}=\tr_{\fW_2}|\psi_{\fW}\rangle\langle\psi_{\fW}|\ee
\be \log\rho_{\fB0}=\log\rho_{\text{in}}=\left(\log\left( 1-\frac 34 \vep\right)-\log\frac\vep 4\right)|\Gamma_{\bar \fA\fA}\rangle\langle\Gamma_{\bar\fA\fA}|+\log\frac\vep 4\text{Id}\ee
\be \tr \rho_{\fB}\log\rho_{\fB0}=\log\frac\vep4+\left(\log\left( 1-\frac 34 \vep\right)-\log\frac\vep 4\right)\langle\Gamma_{\bar\fA\fA}|\rho_{\fB}|\Gamma_{\bar \fA \fA}\rangle=\frac 34\log\vep+O(\vep)\ee
where we used
\be \langle\Gamma_{\bar\fA\fA}|\rho_{\fB}|\Gamma_{\bar \fA \fA}\rangle=\frac 14 \sum_{ij}\langle ii|\text{Id}_{\bar\fA}\otimes\tilde\rho_{\fW}|jj\rangle=\frac 14\,.\ee
Then we have
\be\label{swapJ} J_{\text{swap}}(\fA:\fB)=-\frac 34 \log \vep+\cdots\ee
where the dots stands for terms that are bounded as $\vep\to 0$ and include contributions from the mutual information $I_{\text{swap}}(\fB:\fW)$. We see that the swap (\ref{swapJ}) leads to a smaller $J$ than $V=X_A$ in (\ref{XJ}).

\bibliographystyle{jhep}
\bibliography{refs.bib}

\end{document}